\documentclass[lettersize,journal]{IEEEtran}
\normalsize
\usepackage{cite}
\usepackage{color}
\ifCLASSINFOpdf
  \usepackage[pdftex]{graphicx}
  \graphicspath{{figures/}}
  \DeclareGraphicsExtensions{.eps,.jpeg,.png,.jpg}
\else
\fi
\usepackage[cmex10]{amsmath}
\usepackage{amssymb}
\usepackage{amsthm, bm}
\theoremstyle{remark}
\newtheorem*{remark}{Remark}
\usepackage{algorithm,algorithmic}
\usepackage[c2]{optidef}
\ifCLASSOPTIONcompsoc
  \usepackage[caption=false,font=normalsize,labelfont=sf,textfont=sf]{subfig}
\else
  \usepackage[caption=false,font=footnotesize]{subfig}
\usepackage{multirow}
\usepackage{booktabs}

\begin{document}
\setlength{\textfloatsep}{1\baselineskip plus 0.2\baselineskip minus 0.2\baselineskip}
\title{Deep Unsupervised Learning for Joint Antenna Selection and Hybrid Beamforming}
\author{Zhiyan~Liu,
        Yuwen~Yang,
        Feifei~Gao,~\IEEEmembership{Fellow,~IEEE},
        Ting~Zhou,
        and Hongbing~Ma
\thanks{Manuscript received June 5, 2021; revised November 2, 2021; accepted January 4, 2022. This work was supported in part by the National Natural Science Foundation of China under Grant 61831013, by Tsinghua University Initiative Scientific Research Program, and by Tsinghua University-China Mobile Communications Group Co., Ltd. Joint Institute. The work of T. Zhou was supported in part by the Science and Technology Commission Foundation of Shanghai (No. 20DZ1101200). The associate editor coordinating the review of this article and approving it for publication was L. Schmalen. \emph{(Corresponding author: Feifei Gao.)}}
\thanks{Z. Liu, Y. Yang and F. Gao are with Institute for Artificial Intelligence Tsinghua University (THUAI), State Key Lab of Intelligent Technologies and Systems, Beijing National Research Center for Information Science and Technology (BNRist), Tsinghua University, Beijing 100084, P. R. China (e-mail: zyliu@eee.hku.hk; yyw18@mails.tsinghua.edu.cn; feifeigao@ieee.org).}
\thanks{T. Zhou is with the Shanghai Frontier Innovation Research Institute, Chinese Academy of Sciences, Shanghai 201210, P.R. China (e-mail: zhouting@sari.ac.cn).}
\thanks{H. Ma is with the Department of Electronic Engineering, Tsinghua University, Beijing 100084, P. R. China (e-mail: hbma@mail.tsinghua.edu.cn).}}

\markboth{Accepted by IEEE Transactions on Communications}%
{Liu \MakeLowercase{\textit{et al.}}: Deep Unsupervised Learning for Joint Antenna Selection and Hybrid Beamforming}

\maketitle

\begin{abstract}

In this paper, we propose a novel deep unsupervised learning-based approach that jointly optimizes antenna selection and hybrid beamforming to improve the hardware and spectral efficiencies of massive multiple-input-multiple-output (MIMO) downlink systems. By employing ResNet to extract features from the channel matrices,  two neural networks,  i.e., the antenna selection network (ASNet) and the hybrid beamforming network (BFNet), are respectively proposed for dynamic antenna selection and hybrid beamformer design. 
Furthermore,  a deep probabilistic subsampling trick and a specially designed quantization function are respectively developed for  ASNet and BFNet to  preserve  the differentiability while  embedding  discrete  constraints into the  network structures.
With the aid of a flexibly designed loss function,   ASNet and BFNet are jointly trained in a phased unsupervised way, which 
 avoids the prohibitive computational cost of acquiring training labels in supervised learning.
Simulation results demonstrate the advantage of the proposed approach over conventional optimization-based algorithms in terms of both the achieved rate and the computational complexity.

\end{abstract}


\IEEEpeerreviewmaketitle

\section{Introduction}
\IEEEPARstart{M}{assive} multiple-input-multiple-output (MIMO) is a promising technology that significantly improves the spectral efficiency with a large number of antennas\cite{1}. It is considered as a key technology for the fifth-generation wireless communication systems (5G) and beyond. However, the fully-digital implementation of massive MIMO systems requires connecting every antenna to an independent radio frequency (RF) chain. Consequently, the dramatically increased number of RF chains leads to extra hardware cost and power consumption, and greatly undermines the feasibility of implementing large-scale antenna arrays at base station (BS), especially in millimeter wave (mmWave) frequency bands \cite{7010533}. 

To address this issue, hybrid beamforming has been proposed to preserve most of the array gain with reduced hardware cost and power consumption \cite{hbf1,hbf2}. In hybrid beamforming, the signal is first processed by a low-dimensional digital beamformer and then passes through a high-dimensional analog beamformer constructed with analog phase shifters that have constant modulus. Unfortunately, hybrid beamformer design is generally accompanied with non-convex optimization and is thus a difficult task. A majority of works focus on minimizing the distance between the hybrid beamformers and the fully-digital beamformers while assume infinite-resolution phase shifters \cite{hbf2,moaltmin,hbf3}. However, accurate phase shifters are generally associated with complicated circuits and high energy consumption in practice. Hence, it is typical to use cost-effective phase shifters with low-resolution inputs \cite{7370753}. Naturally, directly applying algorithms \cite{hbf2,moaltmin,hbf3} into low-resolution phase shifters would suffer from severe performance loss. Several hybrid beamforming algorithms have been designed for low-resolution phase shifters. To maximize the spectral efficiency, the authors proposed to iteratively update the columns of the analog beamformer in \cite{sohrabi1,sohrabi2}, while in \cite{jchen}, the authors proposed to successively design the low-resolution analog beamformer and combiner. Meanwhile, to minimize the distance between the hybrid beamformers and the fully-digital beamformers, the authors in \cite{zwang} proposed an iterative algorithm based on the coordinate descent method (CDM) while the authors in \cite{9411813} proposed to apply lattice decoders. However, all the aforementioned works on hybrid beamforming \cite{hbf2,moaltmin,hbf3,sohrabi1,sohrabi2,jchen,zwang,9411813} adopt the fully-connected architecture that still requires a large number of phase shifters. In fact, it is possible to further improve hardware efficiency by introducing antenna selection, where only a subarray is activated. 

In the past decades, antenna selection has been widely investigated and various algorithms have been developed. In general, finding an optimal antenna subarray relies on the exhaustive search or the branch-and-bound (BAB) search\cite{2} that suffer from high computational complexity, especially for massive MIMO systems. To reduce the complexity, many sub-optimal approaches have been proposed, such as convex optimization \cite{3}, dominant-matrix search \cite{4} and greedy search-based antenna selection \cite{5}. Meanwhile, combining antenna selection with hybrid beamforming is a relatively new research topic. In \cite{6}, the authors proposed an iterative algorithm to jointly optimize the hybrid beamformers and combiners as well as the antenna selection matrix to maximize the sum rate in a multi-user (MU) MIMO system. However, the algorithm can only cope with infinite-resolution phase shifters. The authors of \cite{hli} considered low-resolution phase shifters and designed a joint antenna selection and hybrid beamforming algorithm. However, the discussions are restricted to MISO systems, where the user employs only one antenna. 

Recently, leveraging machine learning (ML) to solve challenging problems in wireless communications has drawn growing attention. ML has been successfully applied in channel estimation\cite{7}, power allocation\cite{9}, signal detection\cite{8}, etc., {while artificial neural networks (ANNs) have proved to be particularly effective\cite{8742579}. In \cite{8227766}, the authors train a deep neural network (DNN) to approximate complex algorithms for wireless resource allocation with greatly reduced computational overhead.} In \cite{mlhbf1}, the author applies a convolutional neural network (CNN) for hybrid beamforming, and the fully-digital beamformer serves as the training label. In \cite{mlhbf2}, the authors proposed a CNN-based framework for singular value decomposition (SVD) and hybrid beamforming, which is trained to minimize the distance between the real and estimated low-rank approximation of channel matrices. There have also been attempts to apply reinforcement learning \cite{mlhbf3} and model-driven learning \cite{mlhbf4} for hybrid beamforming. 
In \cite{mlas2,mlas3}, the authors proposed deep learning-based antenna selection for channel extrapolation to minimize the channel estimation error. Meanwhile, most works formulate the capacity-oriented antenna selection problem as a classification problem, where each category represents a candidate antenna subarray. The problem is then solved by supervised learning, such as support vector machine (SVM) \cite{10} and DNN \cite{mlas1}. In \cite{11}, the authors proposed to successively apply two CNNs for joint antenna selection and hybrid beamforming, whose training labels are acquired with the exhaustive antenna subarray search and conventional hybrid beamforming algorithms respectively. The authors of \cite{tvu} proposed a doubly iterative algorithm, in which an inter loop optimizes the beamforming vectors and an outer loop exhaustively tries all antenna subarrays. To reduce the complexity, a DNN is also employed in \cite{tvu} to fit the outcomes of the outer loop. Note that the algorithms in \cite{10,mlas1,11,tvu} employ supervised learning to train the ML models. Naturally, supervised learning requires the optimal subarray and hybrid beamformer serving as training labels, which involves exhaustive search or BAB search and leads to high computational complexity when the system scales up. Alternatively, one could obtain sub-optimal training labels with the existing algorithms, but the ML models cannot outperform the existing algorithms under this circumstance. 

Considering the difficulty of acquiring labels in many scenarios, one could utilize unsupervised learning to train ML models. In \cite{unsup1}, the authors proposed a DNN-based algorithm for pilot power allocation in which the sum mean square error (MSE) is defined as the loss function, and thus avoid the complexity of acquiring the global optimal solution as labels. In \cite{unsup2, unsup4, unsuphb3}, hybrid beamforming algorithms for infinite-resolution phase shifters are designed through unsupervised learning, where the achieved rate or sum rate performance is chosen as the loss function. However, the extension of the aforementioned approaches to hybrid beamforming with 1-bit phase shifters is not straightforward due to the additional discrete phase constraints. Also, unsupervised learning has not yet been used for the antenna selection problem with binary constraints to the best of the authors' knowledge.

In this paper, we introduce a joint antenna selection and hybrid beamforming transmitter architecture with 1-bit phase shifters for the massive MIMO downlink. The proposed architecture provides rich flexibility to trade-off between performance and hardware efficiency. 
We then propose a novel deep unsupervised learning-based approach to solve the joint optimization problem, which consists of two neural networks: the antenna selection network (ASNet) for dynamic antenna selection and the hybrid beamforming network (BFNet) for hybrid beamformer prediction. Both two networks employ ResNet to extract features from the channel matrix. To embed the discrete problem constraints into the network structures while preserving differentiability, we propose a deep probabilistic subsampling trick for ASNet, while for BFNet, we propose a specially designed quantization function. A flexible loss function consisting of the achieved rate and regularizers is introduced, enabling a phased unsupervised training approach to jointly train the two networks. We also present the complexity analysis to compare the proposed approach with conventional algorithms. Simulation results are given to compare the proposed algorithm with conventional and supervised learning-based methods in terms of achieved rate and computation time, which prove the effectiveness and efficiency of the proposed approach.  

The rest of the paper is organized as follows. Section II introduces the proposed joint antenna selection and hybrid beamforming architecture for the massive MIMO downlink. The unsupervised learning architecture design and the training procedure are presented in Section III and Section IV, respectively. The complexity of the proposed approach is analyzed in Section~V and simulation results are presented in Section VI. Finally, the conclusions are given in Section VII. 

\textit{Notations: }Scalar variables are denoted by normal-face letters $x$, while vectors and matrices are denoted by lower and upper case letters, $\mathbf{x}$ and $\mathbf{X}$, respectively. The real part and imaginary part of $x$ is denotes by $\Re\{x\}$ and $\Im\{x\}$ respectively. We define $[\mathbf{X}]_{:,i}$ and $[\mathbf{X}]_{i,j}$ as the $i$-th column and the $(i,j)$-th element of the matrix $\mathbf{X}$, respectively. The notation $|\mathbf{X}|$ denotes the matrix determinant of $\mathbf{X}$ while $|x|$ denotes the absolute value of $x$. Transpose operators and Hermitian operators are denoted by $(\cdot)^T$ and $(\cdot)^H$, respectively. The Frobenius norm is denoted by $\|\cdot\|_\mathrm{F}$.

\section{System Model and Problem Formulation}

\subsection{System Model}

\begin{figure*}[!t]
  \centering
  \includegraphics[width=1\textwidth]{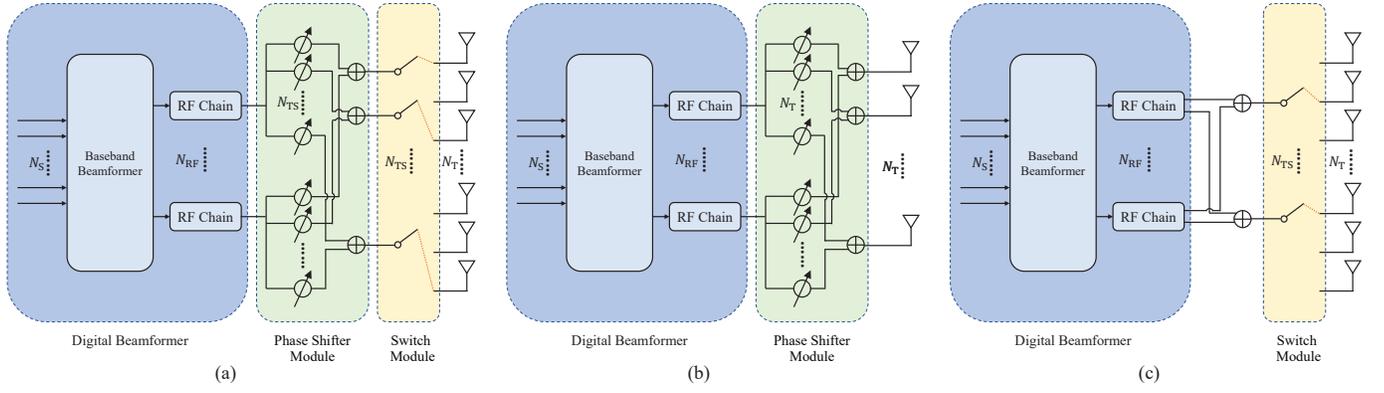}
  \caption{Transmitter architecture for the massive MIMO downlink: (a) Joint antenna selection and hybrid beamforming; (b) Full array with a complicated phase shifter module; (c) Switch-based antenna selection without phase shifters.  }
  \label{fig_system}
  \centering
\end{figure*}

We consider the downlink of a massive MIMO system as shown in Fig.~\ref{fig_system}(a). The BS, equipped with $N_{\mathrm{RF}}$ RF chains and $N_{\mathrm{T}}$ antennas, selects a subarray of $N_{\mathrm{TS}}$ antennas to transmit $N_{\mathrm{S}}$ independent data streams to the receiver with $N_{\mathrm{R}}$ antennas. The numbers of data streams, RF chains and antennas are constrained by $N_{\mathrm{S}} \le N_{\mathrm{RF}} \le N_{\mathrm{TS}} \le N_{\mathrm{T}}$ and $N_{\mathrm{S}} \le N_{\mathrm{R}}$. At the BS, the transmitted symbols are processed by a digital beamformer $\mathbf{T}_{\mathrm{BB}} \in \mathbb{C}^{N_{\mathrm{RF}}\times N_{\mathrm{S}}}$, and then pass through an analog beamformer $\mathbf{T}_{\mathrm{RF}} \in \mathbb{C}^{N_{\mathrm{TS}}\times N_{\mathrm{RF}}}$. To reduce the hardware complexity, we construct the analog beamformer with finite-resolution phase shifters (PSs) that have constant modulus $\frac{1}{\sqrt{N_{\mathrm{TS}}}}$ and $N_\mathrm{B}$-bit inputs. The elements of $\mathbf{T}_{\mathrm{RF}}$ should satisfy $[\mathbf{T}_{\mathrm{RF}}]_{i,j}\in\mathcal{F}$, where $\mathcal{F} = \{\frac{1}{\sqrt{N_{\mathrm{TS}}}}e^{\frac{2\pi b}{2^{N_\mathrm{B}}}}|b=0,1,\ldots ,2^{N_\mathrm{B}}-1\}$. In the scope of this paper, we consider the 1-bit phase shifter case with $N_\mathrm{B}=1$. Meanwhile, the hybrid beamformers should satisfy the power constraint $\|\mathbf{T}_{\mathrm{RF}}\mathbf{T}_{\mathrm{BB}}\|_\mathrm{F}^2 = N_{\mathrm{S}}$. The transmitted signal $\mathbf{x} \in \mathbb{C}^{N_{\mathrm{TS}}}$ can be written as
\begin{equation} \label{signal_x}
\mathbf{x} = \mathbf{T}_{\mathrm{RF}}\mathbf{T}_{\mathrm{BB}}\mathbf{s}, 
\end{equation}
where $\mathbf{s} \in \mathbb{C}^{N_\mathrm{S}}$ is the vector of transmitted Gaussian symbols with $\mathbb{E}\{\mathbf{s}\mathbf{s}^H\} = \mathbf{I}_{N_{\mathrm{S}}}/N_{\mathrm{S}}$.

As shown in Fig.~\ref{fig_system}(a),  we design a fully-connected switching module\footnote{Some works on antenna selection adopt sub-array switching (SAS) structures, where the full array is divided into several disjoint subsets, and only one antenna in each subset can be activated\cite{15}. The proposed approach can be easily extended to SAS structures by adding constraints to the selection matrix. } such that an arbitrary subarray of size $N_{\mathrm{TS}}$ can be activated to transmit $\mathbf{x}$. 
The selection is represented by a binary selection matrix $\mathbf{A}\in \{0,1\}^{N_\mathrm{T}\times N_{\mathrm{TS}}}$ where $[\mathbf{A}]_{m,n}=1$ represents that the $n$-th selected antenna is the $m$-th transmit antenna. Naturally, $\mathbf{A}$ is constrained by $[\mathbf{A}]_{i,j}\in\{0,1\}$ and $\mathbf{A}^T\mathbf{A}=\mathbf{I}$ to ensure that every column is a unit-vector and no duplicate column exists. Then, $\mathbf{H}_\mathrm{S} = \mathbf{HA} \in \mathbb{C}^{N_\mathrm{R} \times N_{\mathrm{TS}}}$ is the channel matrix with the selected transmit antennas, where $\mathbf{H}\in\mathbb{C}^{N_\mathrm{R} \times N_\mathrm{T}}$ denotes the downlink channel matrix with $\|\mathbf{H}\|_\mathrm{F}^2 = N_\mathrm{R}N_\mathrm{T}$. 
Considering block-fading channels, the received symbol $\mathbf{y}\in\mathbb{C}^{N_\mathrm{R}}$ can be written as
\begin{equation} \label{signal_y}
\mathbf{y} = \sqrt{\rho}\mathbf{HA}\mathbf{T}_{\mathrm{RF}}\mathbf{T}_{\mathrm{BB}}\mathbf{s}+\mathbf{n},
\end{equation}
where $\rho$ denotes the average received power, $\mathbf{n}\in\mathbb{C}^{N_\mathrm{R}}$ denotes the additive white complex Gaussian noise with $\mathbf{n}\sim\mathcal{CN}(\mathbf{0},\sigma_n^2\mathbf{I}_{N_\mathrm{R}})$. Note that with $N_{\mathrm{TS}} = N_{\mathrm{T}}$, the transmitter architecture falls back to the conventional hybrid beamforming structure, where all transmitting  antennas are activated and only a fully-connected phase shifter module is involved as shown in Fig.~\ref{fig_system}(b). Although the conventional architecture enjoys a high degree of freedom and can achieve the near-optimal performance, it requires $N_{\mathrm{RF}}N_{\mathrm{T}}$ phase shifters. On the other hand, with $N_{\mathrm{TS}} = N_{\mathrm{RF}}$, the phase shifters can be eliminated and the transmitter architecture is reduced to the switch-based antenna selection architecture as shown in Fig.~\ref{fig_system} (c). However, since $N_{\mathrm{TS}}=N_{\mathrm{RF}}\ll N_{\mathrm{T}}$, the switch-based architecture cannot fully explore the channel diversity of massive MIMO systems, leading to significant performance penalties\cite{7370753}. 
Therefore, We focus on  the  transmitter architecture with $N_{\mathrm{RF}}<N_{\mathrm{TS}}<N_{\mathrm{T}}$ as shown in Fig.~\ref{fig_system}(a). The proposed architecture can provide flexible trade-off between the hardware cost and system performance by jointly exploiting the switch module and the phase shifter module with $N_{\mathrm{RF}}N_{\mathrm{TS}}$ phase shifters. 

\subsection{Channel Model}

The downlink channel is represented by the Saleh-Valenzuela (SV) model. We define $\left| \alpha_l \right|$ as the magnitude of the path gain, $\phi_l$ as the phase of the path gain, $\tau_l$ as the path delay, $B$ as the system bandwidth, $\theta_l^\mathrm{A}$ and $\theta_l^\mathrm{D}$ as the elevation AoA and AoD, $\phi_l^\mathrm{A}$ and $\phi_l^\mathrm{D}$ as the azimuth AoA and AoD of the $l$-th ray, $\mathbf{a}_\mathrm{r}^{*}(\cdot)$ and $\mathbf{a}_\mathrm{t}^{*}(\cdot)$ as the receive and transmit steering vectors, and $\gamma$ as the normalization factor. By assuming uniform linear array (ULA) with the antenna spacing distance $d = \frac{\lambda}{2}$ at both the receiving and transmitting sides\footnote{The ULA model is used here for simpler illustration. In fact, the proposed approach does not restrict to the specifical  array shape and works for arbitrary arrays. }, the channel matrix $\mathbf{H}$ can be written as
\begin{equation} \label{sv-model}
\mathbf{H} = \gamma\sum_{l=1}^{L}\left| \alpha_l \right|e^{j\phi_l}e^{2\pi\tau_lB}\mathbf{a}_\mathrm{r}(\theta_l^\mathrm{A},\phi_l^\mathrm{A})\mathbf{a}_\mathrm{t}^{*}(\theta_l^\mathrm{D},\phi_l^\mathrm{D}).
\end{equation}
with
\begin{align}
\begin{array}{l}
\left\{
\begin{array}{ll} 
\mathbf{a}_\mathrm{r} = [1,e^{j2\pi\sin{(\theta_l^\mathrm{A})}\cos{(\phi_l^\mathrm{A}})},...,e^{j2\pi(N_\mathrm{R}-1)\sin{(\theta_l^\mathrm{A})}\cos{(\phi_l^\mathrm{A}})}]^T,\\
\mathbf{a}_\mathrm{t} = [1,e^{j2\pi\sin{(\theta_l^\mathrm{D})}\cos{(\phi_l^\mathrm{D}})},...,e^{j2\pi(N_\mathrm{T}-1)\sin{(\theta_l^\mathrm{D})}\cos{(\phi_l^\mathrm{D}})}]^T. 
\end{array} \right.
\end{array}
\end{align}

In this work, we assume that $\mathbf{H}$ has been perfectly obtained at the BS side, which is also common for most antenna selection and hybrid beamforming algorithms\cite{3,4,5,6,hli,sohrabi1,sohrabi2,jchen,zwang}. {In reality, however, obtaining CSI can be a major challenge especially with limited RF chains. While we numerically evaluate the achieved performance with imperfect channel estimation in Section~\ref{rateperf}, we refer readers to \cite{pilotpc1, pilotpc2} for end-to-end deep learning designs which directly use the received pilots for precoding. }

\subsection{Problem Formulation}

We choose the achieved rate as the optimization target for the joint design. Specifically, the optimal antenna subarray and hybrid beamformers $(\mathbf{A}^{\mathrm{opt}},\mathbf{T}_{\mathrm{RF}}^{\mathrm{opt}},\mathbf{T}_{\mathrm{BB}}^{\mathrm{opt}})$ are found by maximizing the achieved rate
\begin{equation} \label{rate}
  R = \log_2 |\mathbf{I} + \frac{\rho}{\sigma_n^2 N_{\mathrm{S}}}\mathbf{H}\mathbf{A}\mathbf{T}_{\mathrm{RF}}\mathbf{T}_{\mathrm{BB}}\mathbf{T}_{\mathrm{BB}}^H\mathbf{T}_{\mathrm{RF}}^H\mathbf{A}^H\mathbf{H}^H|. 
\end{equation}
Then, the formulated joint problem can be written as:
\begin{maxi!}
  {\scriptstyle \mathbf{A},\mathbf{T}_{\mathrm{RF}},\mathbf{T}_{\mathrm{BB}}}{R}{\label{P0}}{}
  \addConstraint{[\mathbf{A}]_{i,j}}{\in\{0,1\}}
  \addConstraint{\mathbf{A}^T\mathbf{A}}{=\mathbf{I}}
  \addConstraint{[\mathbf{T}_{\mathrm{RF}}]_{i,j}}{\in\mathcal{F}}
  \addConstraint{\|\mathbf{T}_{\mathrm{RF}}\mathbf{T}_{\mathrm{BB}}\|_\mathrm{F}^2}{= N_{\mathrm{S}}.}
\end{maxi!}

Note that when the selection matrix $\mathbf{A}$ and the analog beamformer $\mathbf{T}_{\mathrm{RF}}$ are determined, the digital beamformer $\mathbf{T}_{\mathrm{BB}}$ that maximize $R$ can be found from the singular value decompsition (SVD) of the effective channel matrix $\mathbf{H}_{\mathrm{eff}} = \mathbf{H}\mathbf{A}\mathbf{T}_{\mathrm{RF}}$\footnote{Here $\mathbf{T}_{\mathrm{BB}}$ is not constrained to be unitary. Hence, the optimal digital beamformer should be the unitary right singular matrix followed by a water-filling power allocation. We desire the proposed neural network to implicitly learn the power allocation part of the digital beamformer. }\cite{hbf2}. Thus, by reducing the optimization variables to $(\mathbf{A},\mathbf{T}_\mathrm{RF})$, one can cast the joint problem as an NP-hard combinatorial problem, whose optimal solution can be obtained through an exhaustive search over all possible combinations of $\mathbf{A}$ and $\mathbf{T}_\mathrm{RF}$. Unfortunately, the exhaustive search requires  prohibitively high complexity due to the  extremly large searching space, i.e.,  ${N_\mathrm{T} \choose N_\mathrm{TS}}2^{N_\mathrm{RF} N_\mathrm{TS} B}$. Therefore, we propose to utilize deep learning to solve the joint problem with lower complexity.

{
\begin{remark}
   The proposed unsupervised approach can  be  extended to other scenarios as long as the desired targets in these scenarios have closed form expressions.
   However, it is still very important to customize neural networks for specific problem attributes.
   For example, in a multi-user (MU) system with $K$ single-antenna users, the sum rate is given as
  \begin{equation}
    R_\mathrm{MU}=\sum_{k=1}^K \log_2(1+\frac{|\mathbf{h}_k^H \mathbf{A} \mathbf{T}_\mathrm{RF} \mathbf{t}_{\mathrm{BB},k}|^2}
    {\sum_{j\neq k}{|\mathbf{h}_k^H \mathbf{A} \mathbf{T}_\mathrm{RF} \mathbf{t}_{\mathrm{BB},j}|^2}+\sigma^2}), 
  \end{equation}
  where $\mathbf{h}_k\in \mathbb{C}^{N_\mathrm{T}}$ denotes the downlink channel from the BS to the $k$-th user, and $\mathbf{t}_{\mathrm{BB},k}$ is the $k$-th column of $\mathbf{T}_\mathrm{BB}$. One can directly transfer the proposed approach by using $\mathbf{H}_\mathrm{MU}=[\mathbf{h}_1, \mathbf{h}_2, \ldots, \mathbf{h}_k]^H$ as the network input and substituting $R_\mathrm{MU}$ for $R$ in the following context. Our numerical results, however, suggest that the proposed approach performs well in the low-SNR region but experiences a degradation in performance in the high-SNR region. The main cause of such degradation is the existence of the trainable parameters in the denominator. In the high-SNR region, the noise variance $\sigma^2$ approaches zero. Hence, as the neural networks are being trained to cancel the inter-user interference, the denominator approaches zero, leading to the gradient explosion problem that greatly hinders the convergence of the neural networks. Possible techniques to resolve such a problem include gradient clipping, target function approximation, or special network designs. These techniques, however, are beyond the scope of this paper.
\end{remark}
}

\section{Network Architecture Design with Unsupervised Learning} \label{sec:3}
Thanks to its excellent  learning  capability and low complexity, deep learning is utilized to 
solve the joint optimization problem (\ref{P0})  by learning the mapping from the channel matrix $\mathbf{H}$ to the optimal subarray and beamformers $(\mathbf{A}^{\mathrm{opt}},\mathbf{T}_{\mathrm{RF}}^{\mathrm{opt}},\mathbf{T}_{\mathrm{BB}}^{\mathrm{opt}})$. 
As a commonly employed deep learning technique, supervised learning  trains neural networks to minimize the distance between the network output and the optimal solution $(\mathbf{A}^{\mathrm{opt}},\mathbf{T}_{\mathrm{RF}}^{\mathrm{opt}},\mathbf{T}_{\mathrm{BB}}^{\mathrm{opt}})$\cite{11,tvu}. However, searching for the optimal solution of the joint problem is infeasible since the problem is NP-hard. Alternatively, sub-optimal labels could be obtained by the existing algorithms to perform supervised training, but will prevent ML models from outperforming the existing algorithms. Hence, we adopt unsupervised learning to circumvent the difficulty of acquiring optimal labels. 
{ Unsupervised learning, although not rigidly defined, generally refers to acquiring knowledge of a data distribution without annotated labels\cite{goodfellow2016deep}. In the context of wireless communication, the propagation scenario $S$ defines a data distribution $p_S(\mathbf{H})$, and unsupervised learning refers to training a set of neural network parameters $\Theta$ to maximize the expectation of the achieved rate over possible channel matrices $\mathbb{E}_{p_S}\{R(\mathbf{H}, \mathrm{Net}(\mathbf{H}, \Theta))\}$ without using annotated data. It is worth noting that the target $\mathbb{E}_{p_S}\{R(\mathbf{H}, \mathrm{Net}(\mathbf{H}, \Theta))\}$ depends on $p_S(\cdot)$, and thus in the learning process the model must implicitly extract knowledge of the data distribution $p_S(\cdot)$. Optimization-based approaches, on the contrary, runs an independent optimization process for each channel matrix without leveraging any learned knowledge of  propagation scenarios. Hence,  unsupervised learning-based approach can potentially achieve better performance than optimization-based approaches with reduced complexity.} Nevertheless, implementing the problem constraints~(6b-6e) is a major challenge in unsupervised learning since the loss function generally implies no constraints on the network output, and thus these constraints must be embedded in the network. Next, we propose a deep learning architecture design consisting of the antenna selection network (ASNet) and the hybrid beamforming network (BFNet) to solve the joint optimization problem (\ref{P0}) in an unsupervised way.

\begin{figure*}[t]
  \centering
  \includegraphics[width=1\textwidth]{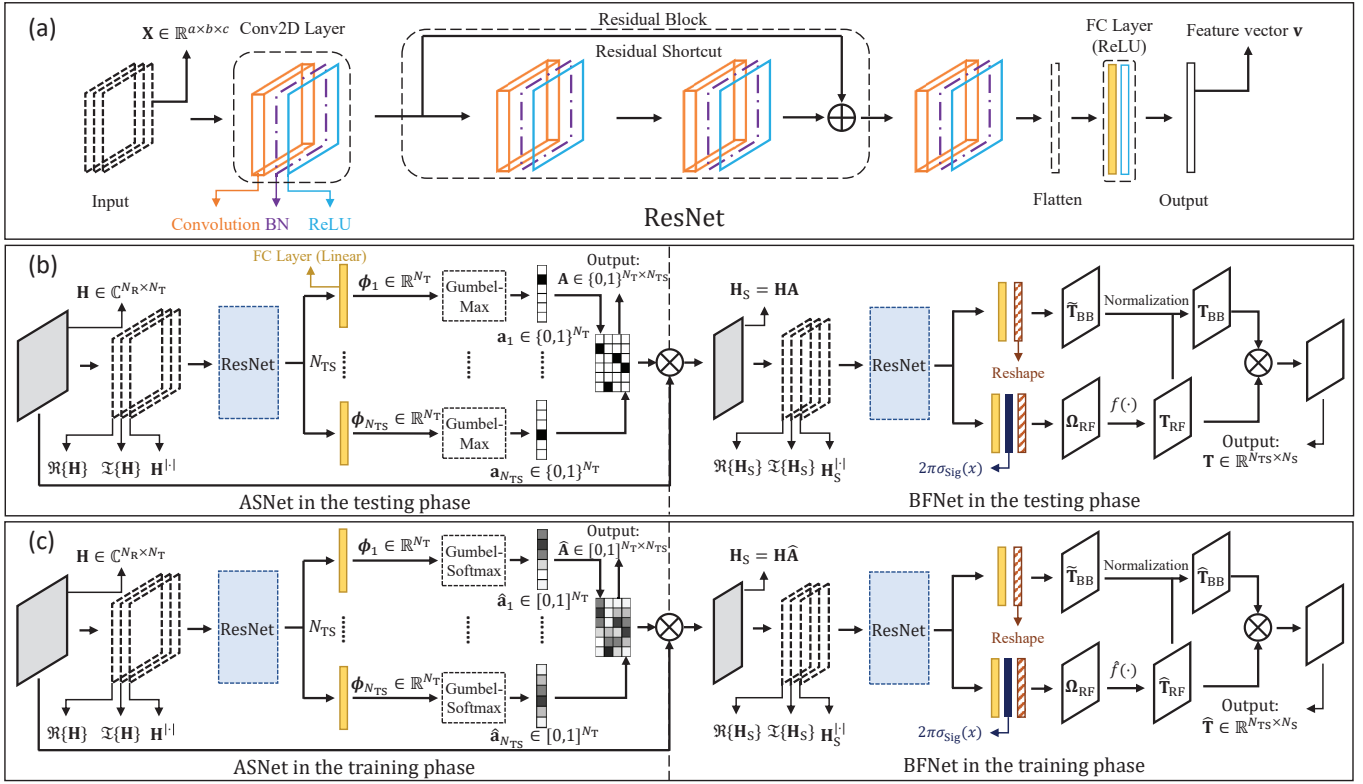}
  \caption{(a) ResNet architecture for feature extraction. (b) The proposed joint network architecture in the testing phase, i.e. $\mathrm{Net}_\mathrm{TEST}$. (c) The proposed joint network architecture in the testing phase, i.e. $\mathrm{Net}_\mathrm{TRAIN}$. }\label{net}
  \centering
\end{figure*}

\subsection{Feature Extraction with ResNet}

The similarity between ASNet and BFNet is the use of deep residual network (ResNet) for extracting features from the channel matrices. {Originally proposed in\cite{resnet}, ResNet exhibits significant superiority over plainly stacked CNNs and has been used as the backbone network for various deep learning tasks. In wireless communications, the effectiveness of ResNet-like architectures has been validated in \cite{mlas3,resnet1,resnet2}.} Define $\mathbf{X}$ as the $a \times b \times c$ real-valued input of ResNet, where $a$, $b$ and $c$ represents the height, width and depth of the input data respectively. In general, ResNet is a cascade of nonlinear transformations of $\mathbf{X}$, which is referred to as layers or blocks. Let us list the details of the elementary layers used here: 
\begin{enumerate}
  \item Conv2D layer: The Conv2D layer is powerful in extracting spatial features of the input data, whose output is given as 
  \begin{equation} \label{conv_layer}
  \mathcal{F}_\mathrm{C}(\mathbf{X}_\mathrm{in}) = \sigma_\mathrm{C}(\mathrm{BN}(\mathbf{W}_\mathrm{C}\ast\mathbf{X}_\mathrm{in}+\mathbf{B}_\mathrm{C})),
  \end{equation}
  where $\mathbf{X}_\mathrm{in}$ denotes the 3-D input tensor, $\ast$ denotes the convolution operation, $\mathbf{W}_\mathrm{C}$ denotes the convolutional filters, $\mathbf{B}_\mathrm{C}$ denotes the bias variable of the convolutional layer, and $\mathrm{BN}(\cdot)$ denotes the batch normalization layer that is widely used in neural networks to stabilize and accelerate the training process\cite{pmlr-v37-ioffe15}. The activation function $\sigma_\mathrm{C}$ for Conv2D layers is the ReLU function, i.e. $\sigma_\mathrm{ReLU}(x)=\max(x,0)$, unless otherwise specified. 
  \item Conv2D residual block: Instead of hoping a few plainly stacked Conv2D layers to fit an underlying mapping, we use Conv2D residual blocks to fit these layers with a residual mapping, which is easier to optimize empirically\cite{resnet}. Unless otherwise specified, the Conv2D residual block consists of two Conv2D blocks and one shortcut, and is defined as 
  \begin{equation} \label{res_block}
  \mathcal{F}_\mathrm{R}(\mathbf{X}_\mathrm{in}) = \mathcal{F}_\mathrm{C}\circ\mathcal{F}_\mathrm{C}(\mathbf{X}_\mathrm{in}) + \mathbf{X}_\mathrm{in}.
  \end{equation}
  \item Fully-connected (FC) layer: The fully-connected layer is a nonlinear transformation from vectors to vectors and can be mathematically represented by 
  \begin{equation} \label{fc_layer}
  \mathcal{F}_\mathrm{FC}(\mathbf{x}_\mathrm{in}) = \sigma_\mathrm{FC}(\mathbf{W}_\mathrm{FC}\mathbf{x}_\mathrm{in}+\mathbf{b}_\mathrm{FC}),
  \end{equation}
  where $\mathbf{x}_\mathrm{in}$ is the 1-D input vector, $\mathbf{W}_\mathrm{FC}$ is the weight variable, $\mathbf{b}_\mathrm{FC}$ is the bias variable, and $\sigma_\mathrm{FC}(\cdot)$ is the element-wise activation function of the fully-connected layer. The possible activation functions include the ReLU function, the sigmoid function $\sigma_\mathrm{Sig}(x)=(e^x-e^{-x})/(e^x+e^{-x})$, and the linear function $\sigma_\mathrm{Lin}(x)=x$.

\end{enumerate}

{The configuration of the layers is determined by trial-and-errors and is illustrated in Fig.~\ref{net}~(a).} Specifically, the input consecutively passes through one Conv2D block, one Conv2D residual block and one more Conv2D block. The resultant feature map is a 3-D tensor, which is then vectorized into a 1-D vector, and fed into one fully-connected layer to generate the final feature vector $\mathbf{v}$ with length $n$. 

\subsection{ASNet: Embedding Constraints for Antenna Selection}\label{ASNet}

ASNet first transforms the input channel matrix $\mathbf{H}$ into the $N_{\mathrm{R}} \times N_{\mathrm{T}} \times 3$ real-valued tensor $\mathbf{X}_\mathrm{AS}$ by stacking the real part, the imaginary part and the element-wise norm of the channel matrix $\mathbf{H}$, i.e. $[\mathbf{X}_\mathrm{AS}]_{m,n,1} = \Re\{[\mathbf{H}]_{m,n}]\}$, $[\mathbf{X}_\mathrm{AS}]_{m,n,2} = \Im\{[\mathbf{H}]_{m,n}]\}$ and $[\mathbf{X}_\mathrm{AS}]_{m,n,3} = \vert[\mathbf{H}]_{m,n}\vert$. Then, the above mentioned ResNet architecture can be applied to extract features from $\mathbf{X}_\mathrm{AS}$ and generate the feature vector $\mathbf{v}_\mathrm{AS}$. {For further use, we define $\pmb{\phi}_1,\pmb{\phi}_2,\ldots,\pmb{\phi}_{N_\mathrm{TS}}$ with $\pmb{\phi}_j \in \mathbb{R}^{N_\mathrm{T}}$, where each $\pmb{\phi}_j$ is the output of an independent fully connected layer with $\mathbf{v}_\mathrm{AS}$ as its input and $\sigma_\mathrm{Lin}(x)$ as its activation function. 

In the testing phase, ASNet is expected to predict the antenna selection matrix $\mathbf{A}$ that satisfies constraints (6b) and (6c) from the feature vector $\mathbf{v}_\mathrm{AS}$. To proceed, we write $\mathbf{A} = [\mathbf{a}_1,\mathbf{a}_2,\ldots,\mathbf{a}_{N_\mathrm{TS}}]$, and thus the constraints (6b) and (6c) equivalently require $\mathbf{a}_k$ to be a unit vector of length $N_\mathrm{T}$ for any $j\in \{1,\ldots,N_\mathrm{TS}\}$ and $\mathbf{a}_j^T\mathbf{a}_k=0$ for any $j\neq k$. Using the probabilistic subsampling theory\cite{huijben2019deep}, each $\mathbf{a}_j$ with $j\in \{1,\ldots,N_\mathrm{TS}\}$ can be derived by one-hot encoding a realization of a categorical random variable $Z_j$. Here $Z_j$ is defined as 
\begin{align}
    Z_j\sim \mathrm{Cat}(N_\mathrm{T},\mathbf{p}_j).
\end{align}
where $\mathrm{Cat}(N_\mathrm{T},\mathbf{p}_j)$ denotes a categorical distribution with $\{1,2,\ldots,N_\mathrm{T}\}$ as its sample space and $p(Z_j=k)=p_{j,k}$ as its probability distribution. By definition, $\mathbf{p}_j\in \mathbb{R}^{N_\mathrm{T}}$ is a normalized probability vector with $p_{j,k}\geq 0$ and $\sum_{k=1}^{N_T}p_{j,k} = 1$. It can be seen that $p_{j,k}$ is simply the probability that the $j$-th selected antenna is the $k$-th transmit antenna. 
To incorporate the random variable $Z_j$ into the
back-propagation algorithm, we first introduce the Gumbel-Max trick \cite{gumbel1954statistical,18,19} to draw discrete realizations from the categorical distribution  $\mathrm{Cat}(N_\mathrm{T},\mathbf{p}_j)$ and then relax the discrete realization by the softmax function to endow it with differentiability.
More specifically, we reparametrize $\mathbf{p}_j$ with the log-probability (logit) vector $\pmb{\phi}_i$ such that
\begin{equation} \label{logits-to-prob}
  p_{j,k} = \frac{\exp \phi_{j,k}}{\displaystyle\sum\limits_{k'=1}^{N_\mathrm{T}}\exp \phi_{j,k'}}.
\end{equation}
With the Gumbel-Max trick,  a realization of $Z_j$ is given by
\begin{equation}
    z_j = \mathop{\arg \max}_{k\neq z_1,\ldots,z_{j-1}}\{\phi_{j,k}+g_{j,k}\},
\end{equation}
where $g_{j,k}$ are i.i.d. samples drawn from $\mathop{\mathrm{Gumbel}}(0,1)$
\footnote{The cumulative distribution function (CDF) of $X\sim\mathop{\mathrm{Gumbel}}(0,1)$ is given as $p(X<x)=\exp(-\exp(-x))$.}, and thus the $j$-th column of $\mathbf{A}$ is
\begin{equation}
    \mathbf{a}_j = \mathrm{onehot}\{z_j\} = \mathrm{onehot}\{\mathop{\arg \max}_{k\neq z_1,\ldots,z_{j-1}}\{\phi_{j,k}+g_{j,k}\}\}. 
\end{equation}
Note that the subscript $k\neq z_1,\ldots,z_{j-1}$ ensures that $\mathbf{a}_j^T\mathbf{a}_k=0$ for any $j\neq k$, e.g. no antenna is selected twice. In this way, we parametrize the binary antenna selection matrix $\mathbf{A}$ with $\pmb{\phi}_1,\ldots,\pmb{\phi}_{N_\mathrm{TS}}$, which are continuous outputs of fully connected layers. }
Next, we approximate the non-differentiable $\mathop{\mathrm{onehot}}\{\mathrm{\mathop{arg\, max}}\{\cdot\}\}$ operator by the continuous softmax function.  
Denote $\hat{\mathbf{A}}=[\hat{\mathbf{a}}_1,\hat{\mathbf{a}}_2,\ldots,\hat{\mathbf{a}}_{N_\mathrm{TS}}]$ as the continuous approximation of ${\mathbf{A}}$, where $\hat{\mathbf{a}}_j \in \mathbb{R}^{N_\mathrm{T}}$ can be element-wisely  written as 
\begin{equation}\label{a_tilde}
  \hat{a}_{j,k} = \frac{\exp ((\phi_{j,k}+g_{j,k})/\tau)}{\displaystyle\sum\limits_{k'=1}^{N_\mathrm{T}}{}\exp ((\phi_{j,k'}+g_{j,k'})/\tau)}.
\end{equation}

Here, $\tau$ represents the softmax temperature. As $\tau\to 0$, the columns of $\hat{\mathbf{A}}$ become exactly one-hot unit vectors. At the beginning of training, we use a high initial temperature $\tau_\mathrm{init}$ to avoid vanishing gradients. Afterwards, we gradually lower the temperature to a small but non-zero value $\tau_\mathrm{final}$ to improve the accuracy of the continuous approximation. {Note that in this approximation we not only relaxed the one-hot constraint but also dropped the orthogonality constraint, i.e. $\mathbf{a}_j^T\mathbf{a}_k=0$ for any $j\neq k$. Hence, we penalize $\sum_{j\neq k} |\mathbf{a}_j^T\mathbf{a}_k|^2$ in the loss function to promote training towards orthogonality, as discussed in Section~\ref{sec-loss}.}

\subsection{BFNet: Embedding Constraints for Hybrid Beamformer Design}\label{BFNet}

The input of BFNet is the channel matrix with selection $\mathbf{H}_\mathrm{S} \in \mathbb{C}^{N_\mathrm{R}\times N_\mathrm{TS}}$.
Similar to ASNet, we transform $\mathbf{H}_\mathrm{S}$ into a ${N_\mathrm{R}\times N_\mathrm{TS}\times 3}$ real-valued tensor $\mathbf{X}_\mathrm{BF}$ with three channels $[\mathbf{X}_\mathrm{BF}]_{m,n,1} = \Re\{[\mathbf{H}_\mathrm{S}]_{m,n}]\}$, $[\mathbf{X}_\mathrm{BF}]_{m,n,2} = \Im\{[\mathbf{H}_\mathrm{S}]_{m,n}]\}$ and $[\mathbf{X}_\mathrm{BF}]_{m,n,3} = \vert[\mathbf{H}_\mathrm{S}]_{m,n}\vert$, and use a ResNet architecture to extract features from $\mathbf{X}_\mathrm{BF}$ and generate the feature vector $\mathbf{v}_\mathrm{BF}$. 

{ To predict the analog beamformer $\mathbf{T}_\mathrm{RF}$, a fully connected layer with the scaled sigmoid function $\sigma'_\mathrm{Sig}(x)=2\pi \sigma_\mathrm{Sig}(x)$ as its activation function is applied to the feature vector $\mathbf{v}_\mathrm{BF}$, whose output is defined as $\omega_\mathrm{RF}\in\mathbb{R}^{N_\mathrm{TS}\times N_\mathrm{RF}}$. Then, $\omega_\mathrm{RF}$ is reshaped into a matrix $\Omega_\mathrm{RF} \in \mathbb{R}^{N_\mathrm{TS}\times N_\mathrm{RF}}$ such that $\mathrm{vec}(\Omega_\mathrm{RF})=\omega_\mathrm{RF}$. Note that the range of $\sigma'_\mathrm{Sig}(x)$ is $[0,2\pi]$. Hence, the elements of  $\Omega_\mathrm{RF}$ fall into the interval $[0,2\pi]$, and are thus defined as the unquantized phase values of the analog beamformer $\mathbf{T}_\mathrm{RF}$.} As the analog beamformer is implemented with 1-bit phase shifters, the elements of $\mathbf{T}_\mathrm{RF}$ should have constant modulus $\frac{1}{\sqrt{N_\mathrm{TS}}}$ and the discrete phase value $0$ or $\pi$. To satisfy the phase constraints, in the testing phase the predicted continuous phase value $\Omega_\mathrm{RF}$ will be quantized using the quantization function
\begin{equation}\label{simple_quant}
  f(\theta) = \pi \left\lfloor \frac{\theta}{\pi} \right\rfloor. 
\end{equation}  
\begin{figure}[t]
  \centering
  \includegraphics[width=75mm]{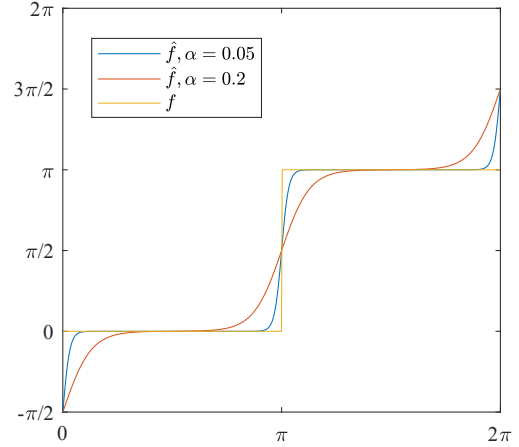}
  \caption{The original quantization function $f$ and the proposed approximating quantization function $\hat{f}$ with $\alpha=0.2$ and $\alpha=0.05$. }\label{fig_quant}
  \centering
\end{figure}
The constrained analog beamformers $\mathbf{T}_\mathrm{RF}$ is then constructed with $[\mathbf{T}_\mathrm{RF}]_{m,n}=\frac{1}{\sqrt{N_\mathrm{TS}}} e^{jf([\Omega_\mathrm{RF}]_{m,n})}$. In the training phase, however, the quantization function $f$ forbids the back-propagation since $f$ has zero gradients almost everywhere. To allow the back-propagation, we approximate $f$ by a function $\hat{f}$ with non-zero gradient in the training phase. Specifically, $\hat{f}$ is written as
\begin{equation}\label{approx_quant}
  \hat{f}(\theta)=
  \begin{cases}
    \frac{\pi}{1+\exp{(-\theta/\alpha})} - \pi& 0\leq x \leq 0.5\pi\\
    \frac{\pi}{1+\exp{(-(\theta-\pi)/\alpha})} & 0.5\pi<x\leq 1.5\pi\\
    \frac{\pi}{1+\exp{(-(\theta-2\pi)/\alpha})} + \pi & 1.5\pi<x\leq 2\pi
  \end{cases},
\end{equation}
where $\alpha$ is a scale factor. As shown in Fig.~\ref{fig_quant}, reducing $\alpha$ improves the approximation accuracy, but can lead to vanishing gradients when $\alpha$ is very close to zero.
With an appropriate value of $\alpha$, the predicted analog beamformer in the training phase will be constructed as $[\hat{\mathbf{T}}_\mathrm{RF}]_{m,n}=\frac{1}{\sqrt{N_\mathrm{TS}}} e^{j\hat{f}([\Omega_\mathrm{RF}]_{m,n})}$.

{To predict the digital beamformer $\mathbf{T}_\mathrm{BB}$, two independent fully connected layer are applied to $\mathbf{v}_\mathrm{BF}$ with $\sigma_\mathrm{Lin}$ as the activation function and $\tilde{\mathbf{t}}_\mathrm{BB}^\mathrm{Re}\in\mathbb{R}^{N_\mathrm{RF}\times N_\mathrm{S}}$ and $\tilde{\mathbf{t}}_\mathrm{BB}^\mathrm{Im}\in\mathbb{R}^{N_\mathrm{RF}\times N_\mathrm{S}}$ as their respective outputs. Then, $\tilde{\mathbf{t}}_\mathrm{BB}^\mathrm{Re}$ is reshaped into a matrix $\tilde{\mathbf{T}}_\mathrm{BB}^\mathrm{Re}\in\mathbb{R}^{N_\mathrm{RF}\times N_\mathrm{S}}$ such that $\mathrm{vec}(\tilde{\mathbf{T}}_\mathrm{BB}^\mathrm{Re})=\tilde{\mathbf{t}}_\mathrm{BB}^\mathrm{Re}$. Similarly, we obtain $\tilde{\mathbf{T}}_\mathrm{BB}^\mathrm{Im}$ by reshaping $\tilde{\mathbf{t}}_\mathrm{BB}^\mathrm{Im}$. The unnormalized digital beamformer $\tilde{\mathbf{T}}_\mathrm{BB}\in \mathbb{C}^{N_\mathrm{RF}\times N_\mathrm{S}}$ is then constructed with $\tilde{\mathbf{T}}_\mathrm{BB} = \tilde{\mathbf{T}}_\mathrm{BB}^\mathrm{Re}+j\tilde{\mathbf{T}}_\mathrm{BB}^\mathrm{Im}$. To satisfy the power constraints, the digital beamformer is normalized in both the testing and training phase, which we give as follows

\begin{align}\label{normalization}
  \mathbf{T}_\mathrm{BB} &= \sqrt{N_\mathrm{S}}\frac{\tilde{\mathbf{T}}_\mathrm{BB}}{\|{\mathbf{T}}_\mathrm{RF}\tilde{\mathbf{T}}_\mathrm{BB}\|_\mathrm{F}},\\   \hat{\mathbf{T}}_\mathrm{BB} &= \sqrt{N_\mathrm{S}}\frac{\tilde{\mathbf{T}}_\mathrm{BB}}{\|\hat{\mathbf{T}}_\mathrm{RF}\tilde{\mathbf{T}}_\mathrm{BB}\|_\mathrm{F}}. 
\end{align}
The resultant hybrid beamformer is then ${\mathbf{T}}={\mathbf{T}}_\mathrm{RF}{\mathbf{T}}_\mathrm{BB}$ for the testing phase and $\hat{\mathbf{T}}=\hat{\mathbf{T}}_\mathrm{RF}\hat{\mathbf{T}}_\mathrm{BB}$ for the training phase. Note that to reduce the computational overhead of SVD and the water-filling algorithm, we stick to the data-driven digital beamformer ${\mathbf{T}}_\mathrm{BB}$ in the testing phase. }

\subsection{The Joint Network}\label{joint_net}

We fuse ASNet and BFNet into a joint network which takes the full channel matrix $\mathbf{H}$ as its input and outputs the predicted selection matrix and hybrid beamformer. The structures of the joint network in the testing stage and the training stage are illustrated in Fig.~\ref{net}~(b) and Fig.~\ref{net}~(c) respectively.  {In the testing stage,} ASNet predicts the selection matrix ${\mathbf{A}}$ from $\mathbf{H}$, based on which the selected channel matrix $\mathbf{H}_\mathrm{S} = \mathbf{H}{\mathbf{A}}$ is generated. Further, $\mathbf{H}_\mathrm{S}$ is fed into BFNet, which predicts the hybrid beamformers ${\mathbf{T}}$. Let $\Theta_\mathrm{AS}$ and $\Theta_\mathrm{BF}$ denote column vectors containing all trainable parameters in ASNet and BFNet, respectively. The joint network can then be compactly written as{
\begin{equation}
  \{{\mathbf{A}}, {\mathbf{T}}\} = \mathrm{Net}_\mathrm{TEST}(\mathbf{H}, \Theta),
\end{equation}
where $\Theta = [\Theta_\mathrm{AS}^T,\Theta_\mathrm{BF}^T]^T$. Both ${\mathbf{A}}$ and ${\mathbf{T}}$ strictly satisfies constraints of the hybrid beamforming problem.}
{With the proposed relaxations, in the training phase the outputs are continuous approximations of ${\mathbf{A}}$ and ${\mathbf{T}}$, which are given as
\begin{equation}
  \{\hat{\mathbf{A}}, \hat{\mathbf{T}}\} = \mathrm{Net}_\mathrm{TRAIN}(\mathbf{H}, \Theta).
\end{equation}}

\section{Training the Joint Network}\label{sec-4}

To train the joint network in an unsupervised way, the loss function $\mathcal{L}$ is a function of the channel matrix $\mathbf{H}$, the predicted selection matrix $\hat{\mathbf{A}}$, and the hybrid beamformer $\hat{\mathbf{T}}$, without requiring the desired optimal solution $(\mathbf{A}^{\mathrm{opt}},\mathbf{T}_{\mathrm{RF}}^{\mathrm{opt}},\mathbf{T}_{\mathrm{BB}}^{\mathrm{opt}})$. Meanwhile, training the joint network is not an easy task because ASNet and BFNet have very different but strongly coupled tasks. Hence, we propose a phased training approach to promote the convergence of the joint network.  

\subsection{Loss Function}\label{sec-loss}

The proposed loss function comprises of two parts: the optimization target and the regularizers. Since the network targets at maximizing the achieved rate (\ref{rate}), we define the optimization target term as
\begin{equation}
  \mathcal{L}_\mathrm{rate} = -\log_2 |\mathbf{I} + \frac{\rho}{\sigma_n^2 N_{\mathrm{S}}}\mathbf{H}\hat{\mathbf{A}}\hat{\mathbf{T}}\hat{\mathbf{T}}^H\hat{\mathbf{A}}^H\mathbf{H}^H|. 
\end{equation}

We add three regularizers to the loss function to accelerate the training and improve the generalization ability. Each regularizer is multiplied by a weight factor $\lambda_i$ for $i = 1,2,3$ to adjust the importance of different penalties. We explain the regularizers as follows:
\begin{enumerate}
  \item To promote training towards an orthogonal selection matrix, we penalize
  \begin{equation}
    \mathcal{L}_\mathrm{pen}^{(1)} = \lambda_1 \sum_{j\neq k} |\hat{\mathbf{a}}_j^T\hat{\mathbf{a}}_k|^2.
  \end{equation}
  where $\hat{\mathbf{a}}_j$ denotes the $j$-th column of the predicted antenna selection matrix $\hat{\mathbf{A}}$.
  \item To accelerate the convergence towards a probability matrix $\mathbf{P}$ with high confidence, we penalize the entropy of the probability distributions as
  \begin{equation}
    \mathcal{L}_\mathrm{pen}^{(2)} = -\lambda_2 \sum_{j=1}^{N_\mathrm{TS}} \sum_{k=1}^{N_\mathrm{T}} p_{j,k} \log_2 p_{j,k}.
  \end{equation}
  \item To prevent the overfitting, we use the well-known $\ell_2$ regularizer on the network parameters\cite{l2}, i.e. 
  \begin{equation}
    \mathcal{L}_\mathrm{pen}^{(3)} = \lambda_3 \|\Theta_\mathrm{TRAIN}\|_2^2, 
  \end{equation}
  where $\Theta_\mathrm{TRAIN}$ denotes a vector containing all trainable network parameters. 
\end{enumerate}

We then formulate the loss function as
\begin{equation}\label{loss}
  \mathcal{L} = \mathcal{L}_\mathrm{rate} + \mathcal{L}_\mathrm{pen}^{(1)} + \mathcal{L}_\mathrm{pen}^{(2)} + \mathcal{L}_\mathrm{pen}^{(3)}. 
\end{equation}
The loss function is averaged over mini-batches with $K$ samples, i.e.
\begin{equation}
  \mathbb{E}\{\mathcal{L}\} = \frac{1}{K}\sum_{k=1}^K\mathcal{L}^{(k)}, 
\end{equation}
where $\mathcal{L}^{(k)}$ denotes the loss function of the $k$-th sample. 
Unless otherwise specified, we apply the adaptive moment estimation (ADAM) algorithm\cite{adam} on $\mathbb{E}\{\mathcal{L}\}$ to update the network parameters. 

\subsection{Phased Training Approach}\label{sec-train}
\begin{algorithm}[!t]
  \caption{Training the joint network}
  \label{algorithm}
  \begin{algorithmic}
  \renewcommand{\algorithmicrequire}{\textbf{Input:}}
  \renewcommand{\algorithmicensure}{\textbf{Output:}}
  \linespread{1.1}\selectfont
  \REQUIRE $N_\mathrm{R}$, $N_\mathrm{T}$, $N_\mathrm{TS}$, $N_\mathrm{RF}$, $N_\mathrm{S}$, $N_1$, $N_2$, $N_3$, batch size $K$, training dataset $\mathcal{H} = \{\mathbf{H}^{(1)}, \mathbf{H}^{(2)},\ldots, \mathbf{H}^{(N)}\}$.
  \ENSURE  $\Theta_\mathrm{AS}$, $\Theta_\mathrm{BF}$
  \\ \textit{Initialize} $\Theta_\mathrm{AS}$, $\Theta_\mathrm{BF}$ and number of batches $n_\mathrm{B}$. 
   \FOR {$n = 1$ to $N_1+N_2+N_3$}
     \FOR {$b = 1$ to $n_\mathrm{B}$}
       \STATE Draw batch $\mathcal{H}_b=\{\mathbf{H}_b^{(1)},\mathbf{H}_b^{(2)},\ldots,\mathbf{H}_b^{(K)}\}$ from $\mathcal{H}$. 
       \STATE $\mathcal{L} \leftarrow 0$.
         \FOR{$k = 1$ to $K$}
           \IF {($1\leq n < N_1$)}
             \STATE Construct $\mathbf{H}_\mathrm{S}^{(k)}$ by selecting $N_\mathrm{TS}$ columns from $\mathbf{H}_b^{(k)}$ randomly.
             \STATE $\hat{\mathbf{T}} \leftarrow \mathop{\mathrm{BFNet}}(\mathbf{H}_\mathrm{S}^{(k)},\Theta_\mathrm{BF})$.
             \STATE $\mathcal{L} \leftarrow \mathcal{L} -\log_2 |\mathbf{I} + \frac{P}{\sigma_n^2 N_{\mathrm{S}}}\mathbf{H}_\mathrm{S}^{(k)}\hat{\mathbf{T}}\hat{\mathbf{T}}^H{\mathbf{H}_\mathrm{S}^{(k)H}}|$.
           \ELSE
             \STATE $\hat{\mathbf{A}} = \mathop{\mathrm{ASNet}}(\mathbf{H}^{(k)}_b,\Theta_\mathrm{AS})$
             \STATE $\hat{\mathbf{T}} \leftarrow \mathop{\mathrm{BFNet}}(\mathbf{H}^{(k)}_b\hat{\mathbf{A}},\Theta_\mathrm{BF})$.
             \STATE $\mathcal{L} \leftarrow \mathcal{L} -\log_2 |\mathbf{I} + \frac{P}{\sigma_n^2 N_{\mathrm{S}}}\mathbf{H}^{(k)}_b\hat{\mathbf{A}}\hat{\mathbf{T}}\hat{\mathbf{T}}^H\hat{\mathbf{A}}^H\mathbf{H}^{(k)H}_b|+\mathcal{L}_\mathrm{pen}^{(1)}+\mathcal{L}_\mathrm{pen}^{(2)}$.
           \ENDIF
         \ENDFOR
         \IF {($1\leq n < N_1$)}
           \STATE $\Theta_\mathrm{TRAIN} \leftarrow \Theta_\mathrm{BF}$, $\mu \leftarrow \mu_1$.
         \ELSIF {($N_1\leq n < N_2$)}
           \STATE $\Theta_\mathrm{TRAIN} \leftarrow \Theta_\mathrm{AS}$, $\mu \leftarrow \mu_2$.
         \ELSE
           \STATE $\Theta_\mathrm{TRAIN} \leftarrow [\Theta_\mathrm{BF}^T, \Theta_\mathrm{AS}^T]^T$, $\mu \leftarrow \mu_3$.
         \ENDIF
         \STATE $\mathcal{L} \leftarrow \mathcal{L}/K + \mathcal{L}_\mathrm{pen}^{(3)}(\Theta_\mathrm{TRAIN})$.
         \STATE Compute $\nabla_{\Theta_\mathrm{TRAIN}} \mathcal{L}$. 
         \STATE $\Theta_\mathrm{TRAIN} \leftarrow \Theta_\mathrm{TRAIN} - \mathop{\mathrm{ADAM}}(\nabla_{\Theta_\mathrm{TRAIN}} \mathcal{L}, \mu)$.
     \ENDFOR
   \ENDFOR
  \RETURN $\{\Theta_\mathrm{BF}, \Theta_\mathrm{AS}\}$.
  \end{algorithmic} 
\end{algorithm}

Since the loss function (\ref{loss}) allows updating the parameters of ASNet and BFNet simultaneously, one could jointly train the two networks from scratches. However, purely conducting joint training is inefficient because: 1) the two networks have very different tasks, thus requiring different learning rates and learning policies; 2) the joint network has a large number of parameters, affecting the convergence speed. Hence, we split the training stage into three phases. In the first phase, we completely exclude ASNet and train BFNet independently with learning rate $\mu_1$ for $N_1$ epochs. The input of BFNet $\mathbf{H}_\mathrm{S}$ is constructed by randomly selecting $N_\mathrm{TS}$ columns of the full channel matrix $\mathbf{H}$. Since $\mathcal{L}_\mathrm{pen}^{(1)}$ and $\mathcal{L}_\mathrm{pen}^{(2)}$ are only related to ASNet, they are excluded from the loss function (\ref{loss}). This design is based on the consideration that BFNet should predict the hybrid beamformer independently, regardless of any specific antenna selection algorithms. In the second phase, the parameters of BFNet are fixed, and we train ASNet to minimize the loss function with learning rate $\mu_2$ for $N_2$ epochs. The second phase trains ASNet to select antenna subarrays that yield the best performance under a specific hybrid beamforming scheme, i.e. BFNet obtained in phase 1. In the third phase, we make all parameters trainable and fine-tune the joint network with a rather small learning rate $\mu_3$ for $N_3$ epochs. Usually, we have $\mu_3 < \mu_1$ and $\mu_3 < \mu_2$. The training algorithm is concluded in Algorithm~\ref{algorithm}.

\section{Complexity Analysis}

We consider the time complexity of the proposed joint network by measuring the number of floating point operations (FLOPS). Consider that the $l$-th convolutional layer has $d_l$ filter kernels of size $h_l^{(1)} \times h_l^{(2)}$ and operates on an $m_l^{(1)}\times m_l^{(2)}$ feature map which has the same shape as the input channel matrix. We define $d_0=3$ as the depth of the input feature map. Then, the complexity of convolutional layers is $\mathcal{O}(\sum_{l\in \mathcal{L}_\mathrm{Conv}}h_l^{(1)}h_l^{(2)}m_l^{(1)}m_l^{(2)}d_{l-1}d_{l})$, where $\mathcal{L}_\mathrm{Conv}$ denotes the index set of all convolutional layers. Since the same kernel size and filter number is used, we have $h_l^{(1)} = h_l^{(2)} = h$ and $d_l = d$ for $l \in \mathcal{L}_\mathrm{Conv}$. As for the dense layers, the complexity is given as $\mathcal{O}(\sum_{l\in \mathcal{L}_\mathrm{FC}}n^\mathrm{in}_{l}n^\mathrm{out}_{l})$, where $\mathcal{L}_\mathrm{FC}$ denotes the index set of all fully-connected layers, and $n^\mathrm{in}_{l}$ ($n^\mathrm{out}_{l}$)  denotes the input (output) data size of the $l$-th layer. The number of neurons in each fully-connected layer is $n$ except for the input and output layers. {Adding up the complexity of each layer, the FLOPS of ASNet is $h^2 d(3d+3)N_\mathrm{T}N_\mathrm{R} + dnN_\mathrm{T}N_\mathrm{R} + nN_\mathrm{T}N_\mathrm{TS}$, and the FLOPS of BFNet is $h^2 d(3d+3)N_\mathrm{TS}N_\mathrm{R} + dnN_\mathrm{TS}N_\mathrm{R} + nN_\mathrm{RF}N_\mathrm{TS} + 2nN_\mathrm{RF}N_\mathrm{S}$. As the system scales up, the kernel size $h$ would remain invariant but $d$ and $n$ would be increased to maintain performance. Hence, considering $N_\mathrm{S} \ll N_\mathrm{T}$ and $N_\mathrm{RF} \ll N_\mathrm{T}$, the complexity of the proposed networks is approximately $\mathcal{O}(d(d+n)N_\mathrm{R}(N_\mathrm{T}+N_\mathrm{TS}))$. 

\begin{table}[!t]
  \renewcommand{\arraystretch}{1.3}
  \caption{Computational Complexity for Antenna Selection and Hybrid Beamforming Algorithms}
  \label{table1}
  \centering
  \begin{tabular}{|c||c|}
  \hline
  Antenna Selection Algorithm     & Computational Complexity                             \\
  \hline
  Exhaustive search\cite{2}          & $\mathcal{O}(N_\mathrm{R}N_\mathrm{T}{N_\mathrm{T} \choose N_\mathrm{TS}})$ \\
  Greedy search-based\cite{5} & $\mathcal{O}(N_\mathrm{R}N_\mathrm{T}N_\mathrm{TS})$ \\
  Convex optimization-based\cite{3} & $\mathcal{O}(N_\mathrm{TS}^3)$ \\
  Proposed ASNet    & {$\mathcal{O}(d(d+n)N_\mathrm{T}N_\mathrm{R})$} \\
  \hline

  \multicolumn{2}{c}{\vspace{0cm}}\\
  \hline
  Hybrid Beamforming Algorithm     & Computational Complexity                             \\
  \hline
  Exhaustive search\cite{jchen}           & $\mathcal{O}(2^{BN_\mathrm{TS}N_\mathrm{RF}}N_\mathrm{TS}^2N_\mathrm{S}N_\mathrm{RF})$ \\
  MO-AltMin\cite{moaltmin}     & $\mathcal{O}(N_\mathrm{TS}^2N_\mathrm{S}N_\mathrm{RF}T)$ \\
  CDM-AltMin\cite{jchen}        & $\mathcal{O}(N_\mathrm{TS}^3N_\mathrm{S}N_\mathrm{RF}^2 T)$ \\
  Babai-AltMin\cite{9411813} & $\mathcal{O}(N_\mathrm{TS}N_\mathrm{S}N_\mathrm{RF}^2 T)$ \\
  Proposed BFNet    & {$\mathcal{O}(d(d+n)N_\mathrm{T}N_\mathrm{TS})$} \\
  \hline
  \end{tabular}
\end{table}

This complexity applies exactly to processing one channel sample in the deployment stage since the forward propagation is executed only one time. The complexity of offline training mainly comes from the gradient computation using back-propagation, which has the same order of complexity as the forward propagation according to\cite{cnntimecost}. Then, the complexity of the entire training process is $\mathcal{O}(N_\mathrm{data}ed(d+n)N_\mathrm{R}(N_\mathrm{T}+N_\mathrm{TS}))$, where $N_\mathrm{data}$ is the number of training samples and $e$ is the number of training epochs. Although it is hard to derive an accurate analytical expression for $e$ since it depends heavily on the training strategy and the data distribution, $e$ is generally polynomial in the number of network layers and training samples according to \cite{dnnconv}. }

We present the complexity of the proposed approach and conventional algorithms for antenna selection or hybrid beamforming in Tab.~\ref{table1}, where $T$ denotes the number of iterations for iterative algorithms. It can be seen that the complexity of conventional algorithms increase faster with respect to $N_\mathrm{R}$, $N_\mathrm{T}$ and $N_\mathrm{TS}$ compared to the proposed approach.  

\section{Numerical Experiments}
In this section, we provide numerical results to demonstrate the performance of the proposed deep unsupervised learning-based approach for antenna selection and hybrid beamforming.
\subsection{Simulation Settings} \label{sim-settings}
\begin{figure}[!t]
  \centering
  \includegraphics[width=80mm]{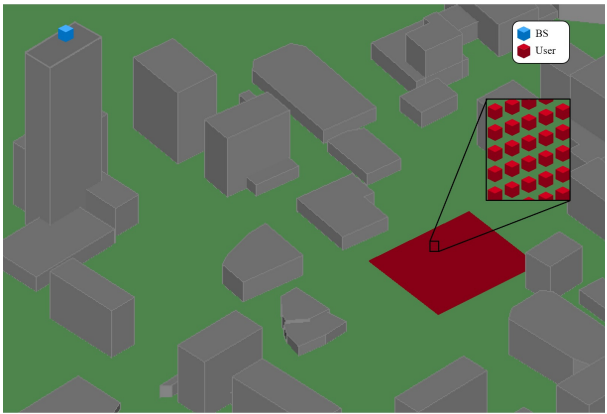}
  \caption{A partial view of the Rosslyn city model adopted in ray-tracing. The blue little box represents the BS, while the red little box represents the possible user locations. Wireless InSite, the ray-tracing simulator, shoots thousands of rays in all directions from the BS and records the strongest 5 paths that reach the user. }\label{fig_city}
  \centering
\end{figure}
We generate the channels for training and evaluating with Wireless InSite, an accurate 3D ray-tracing simulator\cite{wi}. As shown in Fig.~\ref{fig_city}, the Rosslyn city model is adopted as the propagation environment, with a total area of 500$\times$500 $\mathrm{m}^2$. We uniformly select $100,000$ user locations within the red square area, and generate the corresponding downlink channel parameters with the simulator. The simulator considers the $5$ most significant propagation paths connecting the BS and the user at the downlink frequency of $2.5\, \text{GHz}$. Then, we construct the dataset of $100,000$ channel matrices with equation (\ref{sv-model}). Among the data points, we randomly select $70,000$ points as the training samples, while the rest are selected as the testing samples. 
\begin{table}[!t]
  \renewcommand{\arraystretch}{1.3}
  \caption{The Parameters for Training the Neural Networks}
  \label{table2}
  \centering
  \begin{tabular}{ccccc}
    \cline{1-2} \cline{4-5}
    Parameter            & Value       &  & Parameter  & Value     \\ \cline{1-2} \cline{4-5} 
    $\mu_1$               & $10^{-4}$   &  & $\alpha$    & $0.01$    \\
    $\mu_2$               & $10^{-4}$   &  & Batch size  & $512$     \\
    $\mu_3$               & $5\times 10^{-5}$ &  & $\lambda_1$ & $10^{-2}$ \\
    $\tau_\mathrm{init}$  & $1.0$       &  & $\lambda_2$ & $10^{-3}$ \\
    $\tau_\mathrm{final}$ & $0.1$       &  & $\lambda_3$ & $10^{-3}$ \\ \cline{1-2} \cline{4-5} 
  \end{tabular}
\end{table}

All the Conv2D layers used for simulations have 64 filters of size $3\times 3$, while all hidden fully-connected layers have 500 neurons, unless otherwise specified. The number of nodes of the input and output layers correspond with the dimensions of input and output vectors, respectively. The hyper parameters of the training process is listed in Tab.~\ref{table2}. The softmax temperature $\tau$ exponentially decays from $\tau_\mathrm{init}$ to $\tau_\mathrm{final}$ when training ASNet. The models are constructed with TensorFlow 2.4. 
\begin{figure}[!t]
  \centering
  \includegraphics[width=85mm]{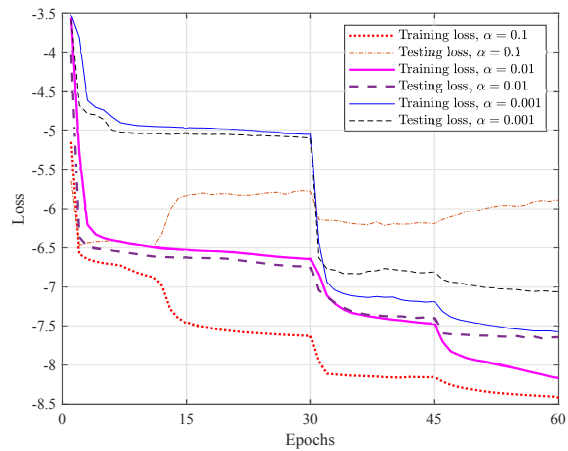}
  \caption{Training and testing loss versus epochs with different scale factors. The network is trained in three phases: (1) Epoch 1-30: training BFNet solely; (2) Epoch 31-45: fixing BFNet and training ASNet; (3) Epoch 46-60: joint training. }\label{fig_train_curve}
  \centering
\end{figure}

In Fig.~\ref{fig_train_curve} we present the training and testing loss versus training epochs with three different scale factors in the approximating quantization function $\hat{f}$: (1) $\alpha=0.1$; (2) $\alpha=0.01$; (3) $\alpha=0.001$. We train BFNet solely with ASNet replaced by random antenna selection with learning rate $\mu_1=10^{-4}$ in the first 30 epochs. In the next 15 epochs, we fix the weights of BFNet and train ASNet with learning rate $\mu_2=10^{-4}$. In the last 15 epochs, the two networks are trained jointly with learning rate $\mu_3=5\times10^{-5}$. From Fig.~\ref{fig_train_curve}, we can immediately see the performance gain after the 30-th epoch when we start to optimize ASNet. Further improvements can be witnessed after the 45-th epoch when joint training starts. Meanwhile, the significant impact of $\alpha$ to the convergence performance is demonstrated. With a relatively large $\alpha=0.1$, the training loss decreases quickly while the testing loss fluctuates, indicating the high mismatch between the approximating quantization function $\hat{f}$ and the real quantization function $f$. On the other hand, $\alpha=0.001$ results in slow convergence and a loss plateau due to vanishing gradients. Hence, using an appropriate $\alpha$ is critical. In particular, the network has the best overall performance with $\alpha=0.01$.

\subsection{Achieved Rate Performance}\label{rateperf}

\begin{figure*}[!t]
\centering
\subfloat[]{\includegraphics[width=0.5\textwidth]{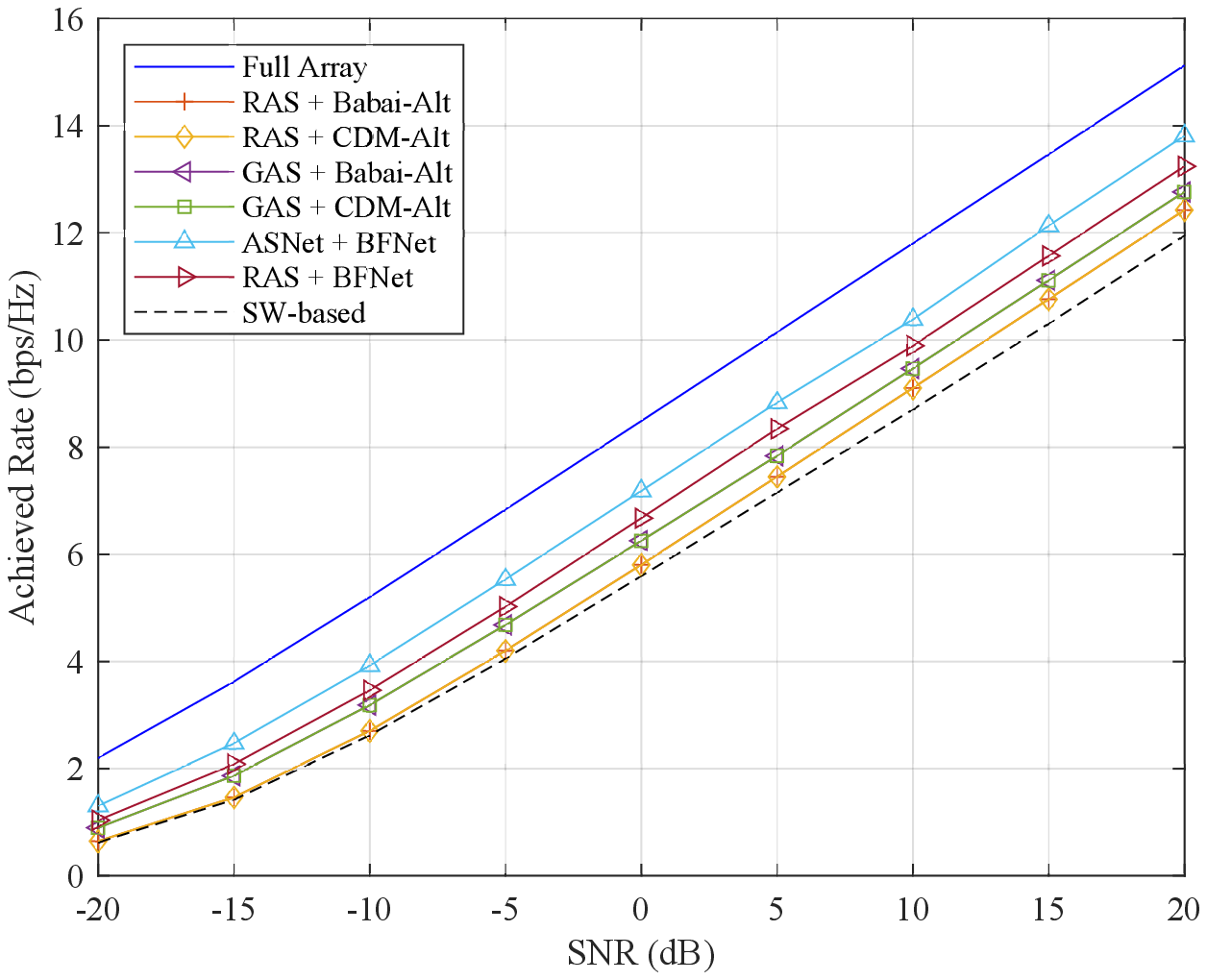}%
\label{fig_ns_1}}
\hfil
\subfloat[]{\includegraphics[width=0.5\textwidth]{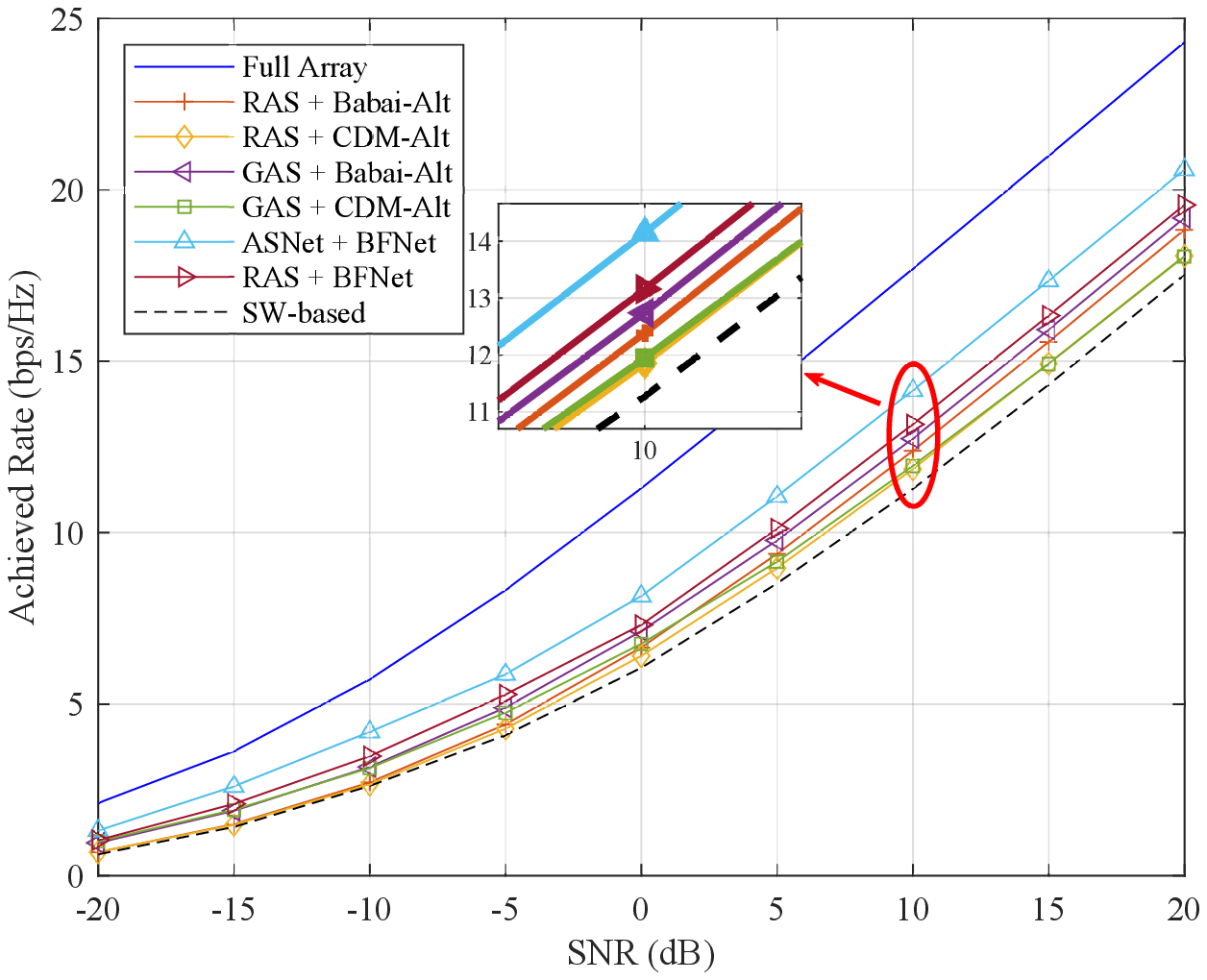}%
\label{fig_ns_2}}
\caption{Achieved rate performance of the proposed approach and conventional algorithms versus SNR for $N_\mathrm{T} = 128$, $N_\mathrm{TS} = 16$, $N_\mathrm{R} = 4$; and (a) $N_\mathrm{RF} = N_\mathrm{S} = 1$ (b) $N_\mathrm{RF} = N_\mathrm{S} = 2$. }
\label{fig_rate_vs_snr}
\end{figure*}

\begin{figure}[!t]
\centering 
\begin{minipage}[b]{0.47\textwidth} 
  \centering
  \includegraphics[width=1\textwidth]{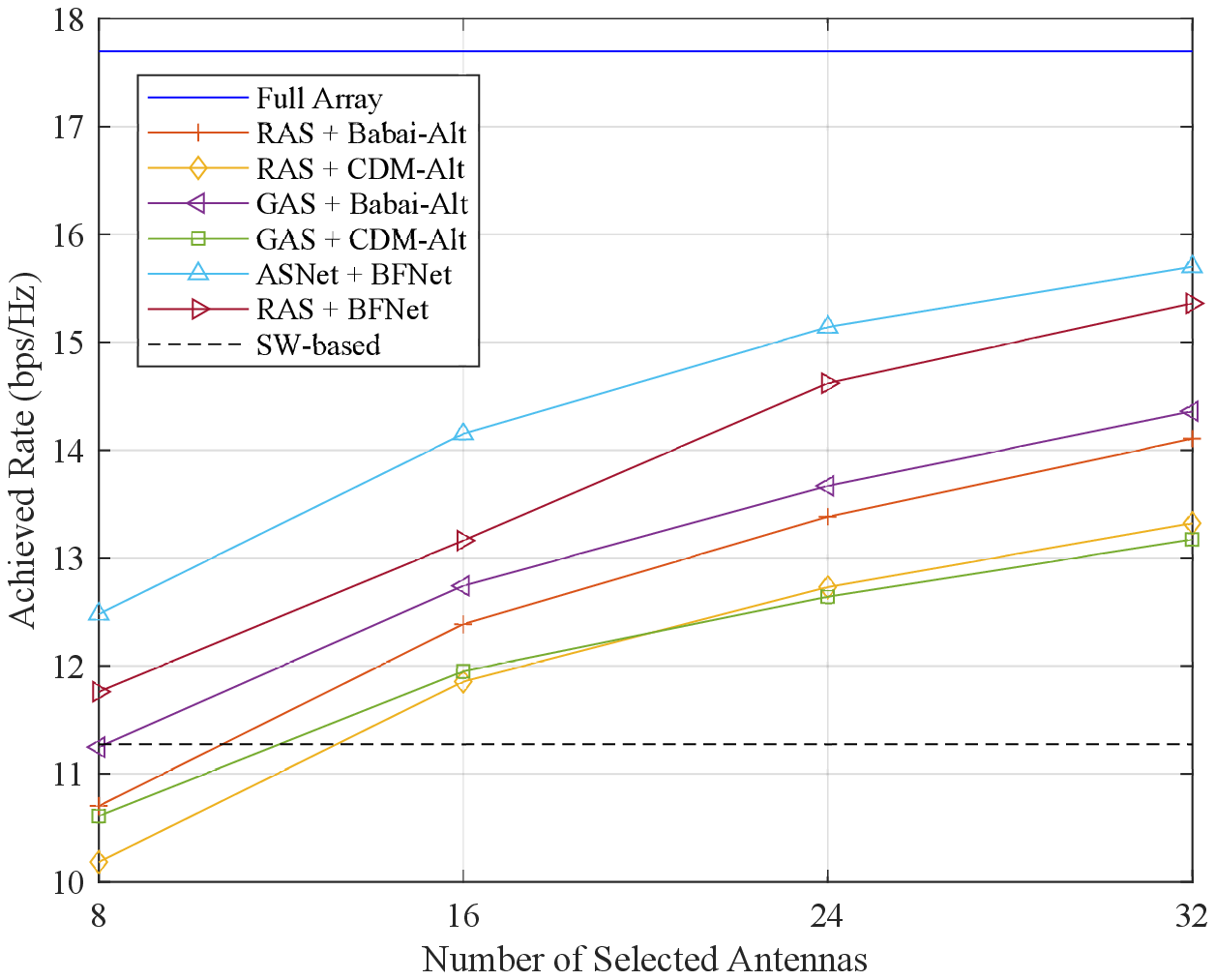}
  \caption{Achieved rate performance of the proposed approach and conventional algorithms versus $N_\mathrm{TS}$ with $N_\mathrm{RF} = N_\mathrm{S} = 2$. }\label{fig_rate_vs_nts}
\end{minipage}
\begin{minipage}[b]{0.47\textwidth} 
  \centering
  \includegraphics[width=1\textwidth]{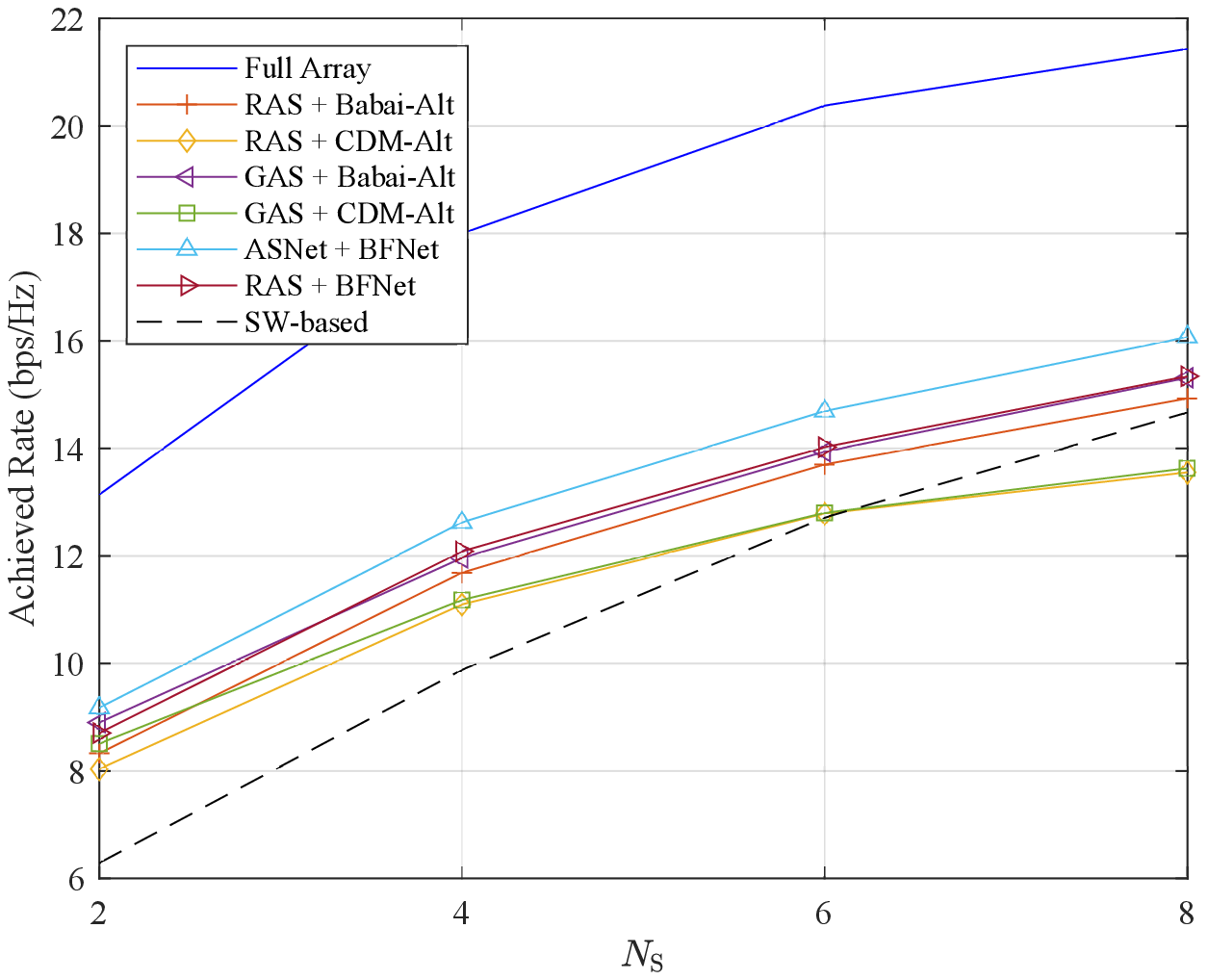}
  \caption{Achieved rate performance of the proposed approach and conventional algorithms versus $N_\mathrm{S}$ with $N_\mathrm{TS} = 16$. }\label{fig_rate_vs_ns}
\end{minipage}
\end{figure}

We first consider the achieved rate performance versus SNR of the proposed deep unsupervised learning-based approach (BFNet + ASNet) and conventional algorithms. In our setting, the BS, equipped with $N_\mathrm{T} = 128$ antennas RF chains communicates with one user with $N_\mathrm{R} = 4$ antennas. In Fig.~\ref{fig_rate_vs_snr}, we present the results for (a) $N_\mathrm{RF} = N_\mathrm{S} = 1$ and (b) $N_\mathrm{RF} = N_\mathrm{S} = 2$ when $N_\mathrm{TS} = 16$ transmitted antennas are selected at the BS side. To the best of the authors' knowledge, there does not exist a conventional algorithm that jointly considers the antenna selection and hybrid beamforming problem. Hence, the baseline algorithms tackle the joint problem in a successive manner: an antenna selection algorithm selects the antenna subarray based on the full channel matrix, and the resultant selected channel matrix is then processed by a hybrid beamforming algorithm. We consider two antenna selection algorithms: the sub-optimal greedy search-based antenna selection algorithm (GAS)\cite{5} and the random antenna selection (RAS)\footnote{Optimal algorithms such as exhaustive search and BAB search would be computationally infeasible at such a system scale.}. As for hybrid beamforming, Babai-Alt\cite{9411813} and CDM-Alt\cite{jchen} are evaluated. To demonstrate the hybrid beamforming performance of BFNet solely, we also provide the achieved rate of BFNet when the antennas are randomly selected. The performance of the full array structure from Fig.~\ref{fig_system}(b) is presented as ``Full Array", while the switch-based antenna selection structure from Fig.~\ref{fig_system}(c) is denoted by ``SW-based". We can see that the proposed approach outperforms all conventional algorithms, which can be attributed to the joint unsupervised training procedure and the well-designed network architecture to efficiently extract features from the channel matrix. Even when the antennas are selected randomly, the hybrid beamforming performance of BFNet is better than the state-of-the-art hybrid beamforming algorithms. It is notable that supervised learning-based hybrid beamforming will not outperform the state-of-the-art algorithm since the latter is used as the ground truth. Thus, the proposed approach is more powerful in hybrid beamformer design than both conventional algorithms and supervised learning-based approach. Compared to the full array architecture, the proposed system design achieves up to $91.3\%$ performance while saving $87.5\%$ number of phase shifters. In Fig.~\ref{fig_rate_vs_nts}, we present the achieved rate performance versus the number of selected antennas with $\text{SNR}=10\, \text{dB}$ and $N_\mathrm{RF} = N_\mathrm{S} = 2$. It can be seen that the proposed algorithm provides the best performance over a wide range of $N_\mathrm{TS}$. Meanwhile, the trade-off between the performance and the hardware complexity can be achieved by varying $N_\mathrm{TS}$. 

{Fig.~\ref{fig_rate_vs_ns} presents the achieved rate performance of the proposed approach with varying $N_\mathrm{S}$ and fixed $N_\mathrm{TS}=16$. To satisfy $N_\mathrm{S}\leq N_\mathrm{RF}$ and $N_\mathrm{S}\leq N_\mathrm{R}$, we set $N_\mathrm{S} = N_\mathrm{RF} = N_\mathrm{R}$. As we can see, the proposed approach consistently outperforms conventional algorithms as $N_\mathrm{S}$, $N_\mathrm{R}$ and $N_\mathrm{RF}$ increase simultaneously. }

\begin{figure}[!t]
  \centering 
  \begin{minipage}[b]{0.47\textwidth} 
    \centering
    \includegraphics[width=1\textwidth]{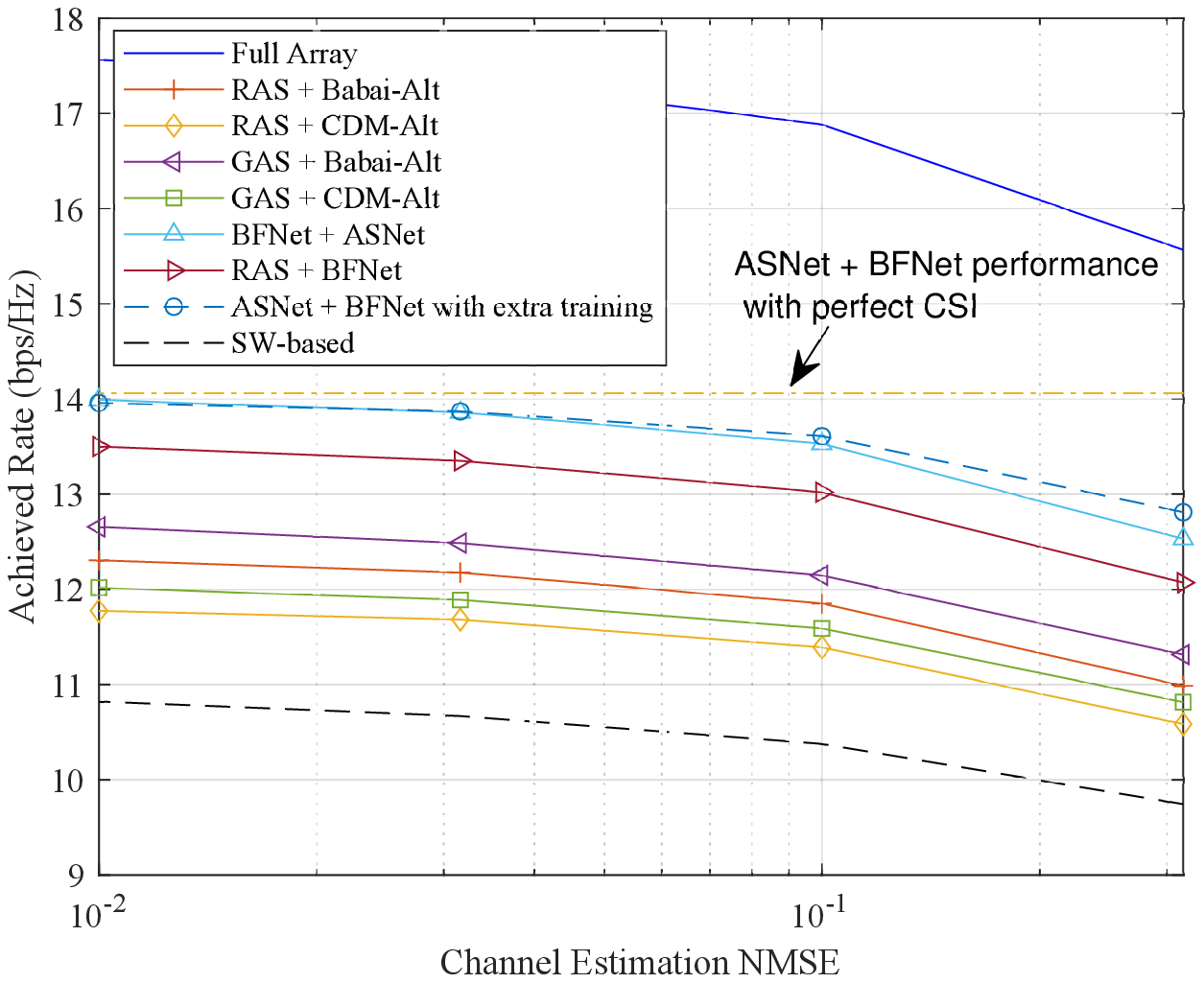}
    \caption{Achieved rate performance of the proposed approach and conventional algorithms versus the channel estimation NMSE. }\label{fig_noise}
  \end{minipage}
  \begin{minipage}[b]{0.47\textwidth} 
    \centering
    \includegraphics[width=1\textwidth]{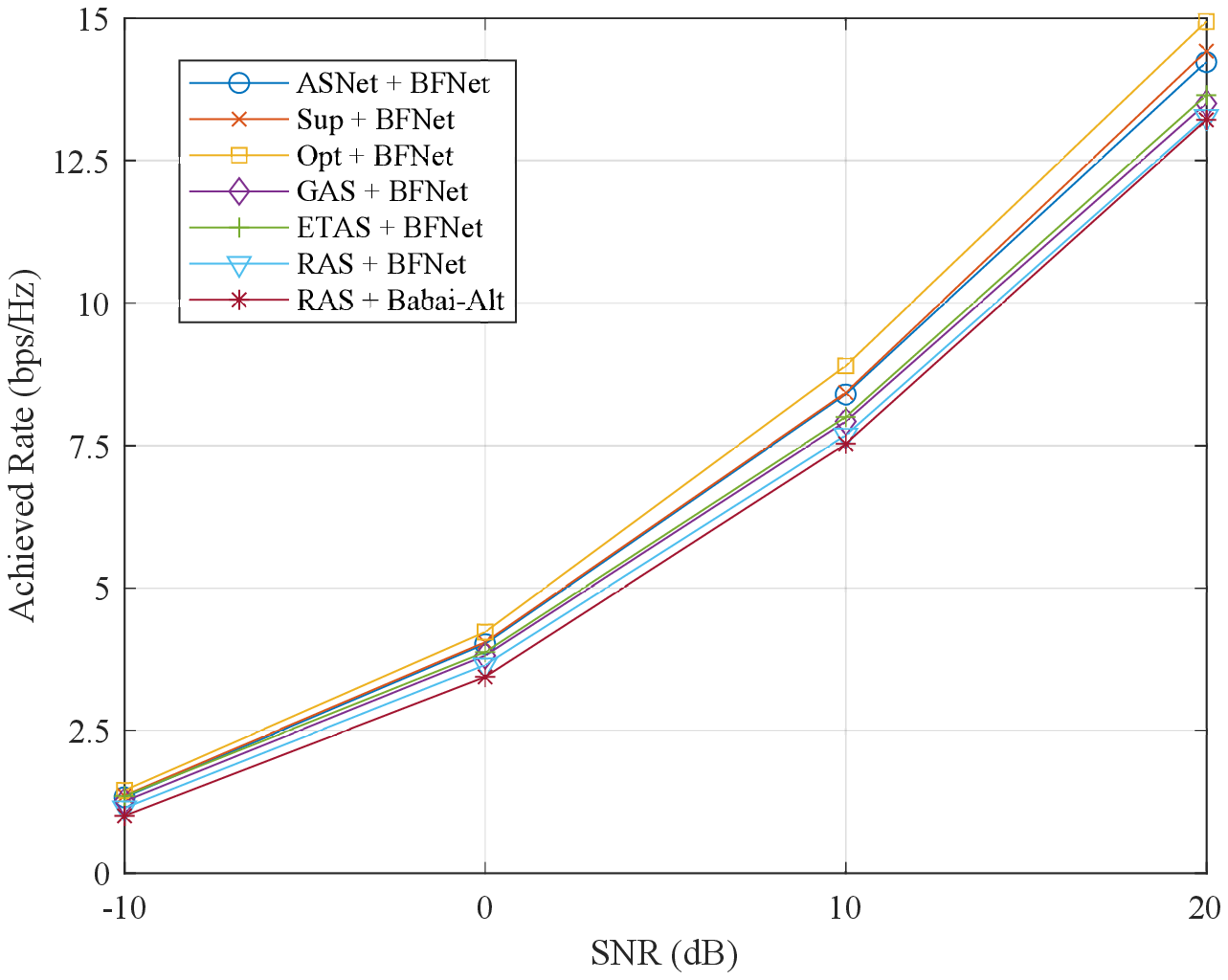}
    \caption{{Achieved rate performance of the proposed approach and other antenna selection schemes versus SNR. }}\label{fig_sup}
  \end{minipage}
\end{figure}

In massive MIMO communications, channel estimation is generally a very challenging task, and demands antenna selection and hybrid beamforming algorithms to be robust to imperfect CSI. Hence, we next evaluate the performance of different algorithms with imperfect channel matrices. { To model the channel estimation error, we consider a time division duplex (TDD) system where the user sends uplink pilots and the BS utilizes the LMMSE channel estimator \cite{lmmse} to estimate the uplink CSI\footnote{The LMMSE channel estimator requires the channel correlation matrix $\mathbf{R}_\mathbf{h}$ to be known at the BS. In our simulation, $\mathbf{R}_\mathbf{h}$ is estimated using the training samples. }. The estimated downlink CSI is thus obtained through channel reciprocity. To improve the robustness of the proposed approach, we use the estimated channel matrices to train the joint network for $3$ more epochs after the joint network has converged on the original dataset. In Fig.~\ref{fig_noise}, we present the achieved rate performance versus the normalized mean square error (NMSE) of channel estimation with $N_\mathrm{RF} = N_\mathrm{S} = 2$. We can see that the proposed approach still performs well when the channel estimation error exceeds $10\%$, and the extra training using corrupted data leads to a performance gain when the NMSE is high. This can be attributed to the fact that the data-driven algorithm learns not only the optimization process but also the data distribution from the highly correlated channel samples. }

{To compare the performance of the proposed unsupervised learning-based antenna selection against optimal antenna selection and supervised learning-based antenna selection, we choose $N_\mathrm{T} = 32$, $N_\mathrm{TS} = 4$, $N_\mathrm{S} = N_\mathrm{RF} = 2$ and $N_\mathrm{R} = 4$. In this case, the number of candidate subarrays is ${32 \choose 4}=35960$ and it takes approximately 160 hours to annotate the entire dataset for supervised learning. The achieved rate performance versus SNR is presented in Fig.~\ref{fig_sup}, where ``Sup'' denotes supervised learning-based antenna selection, ``Opt'' denotes optimal antenna selection, and ``ETAS'' denotes the efficient transmit antenna selection proposed in \cite{4411574}. We can see the proposed approach can significantly reduce the training overhead while maintaining minuscule performance losses compared to supervised learning.}

\subsection{Complexity Performance}\label{cmpperf}

{We numerically evaluate the complexity performance of the proposed approach by measuring the execution time in the training stage and deployment stage respectively, which are compared with supervised learning-based and conventional approaches. For fair comparisons, we implement all the algorithms with MATLAB on the same CPU-based platform, with a 6-core Intel(R) Core(TM) i5 10505 @ 3.20 GHz CPU and 16 GB system memory. 

\subsubsection{Complexity of the training stage}
\begin{table}[!t]
\renewcommand{\arraystretch}{1.3}
\caption{Execution Time of the Training Stage (In Seconds)}
\label{tabel_train}
\centering
\begin{tabular}{@{}ccc@{}}
\toprule
Algorithm           & Dataset Generation & Model Training \\ \midrule
Proposed            & $127.5$              & $4097$         \\
Supervised Learning & $2.292\times 10^5$  & $1139$         \\ \bottomrule
\end{tabular}
\end{table}
Tab.~\ref{tabel_train} summarizes the execution time of the training stage of the proposed and supervised learning-based approach with $N_\mathrm{T} = 32$, $N_\mathrm{TS} = 4$, $N_\mathrm{S} = N_\mathrm{RF} = 2$ and $N_\mathrm{R} = 4$. In our experiments, unsupervised learning needs more training epochs to converge than supervised learning, and thus the model training time for unsupervised learning is longer. However, as we can see, supervised learning requires a very long time to generate the training dataset since the exhaustive search is needed to annotate the channel samples. This computational overhead will grow exponentially with the system scale. On the contrary, the proposed approach completely eliminates the overhead of data annotation and thus requires significantly lower dataset generation time. 

\subsubsection{Complexity of the deployment stage}

\begin{table}[!t]
  \renewcommand{\arraystretch}{1.3}
  \caption{Inference Time of Antenna Selection and Hybrid Beamforming Algorithms (In Seconds)}
  \label{table3}
  \centering
  \begin{tabular}{ccccccc}
  \hline
  $N_\mathrm{T}$  & \multicolumn{3}{c}{128} & \multicolumn{3}{c}{256} \\ \hline
  $N_\mathrm{TS}$ & 8      & 16     & 32     & 16     & 32     & 64     \\ \hline
  Proposed ASNet        & 0.181  & 0.204  & 0.211  & 0.508  & 0.517  & 0.535  \\
  Proposed BFNet        & 0.019  & 0.028  & 0.052  & 0.023  & 0.057  & 0.167  \\
  GAS              & 0.147  & 0.281  & 0.493  & 0.328  & 0.583  & 1.142  \\
  CDM-Alt         & 1.604  & 2.997  & 5.883  & 3.043  & 5.875  & 11.69 \\
  Babai-Alt         & 0.310  & 0.546  & 1.001  & 0.579  & 1.193  & 1.935  \\ \hline
  \end{tabular}
\end{table}

In the deployment stage, the inference time of the proposed approach and other algorithms over $512$ channel samples is measured. For the proposed neural networks, we set the batch size equal to $1$ to model the real-time deployment. The inference time versus $N_\mathrm{T}$ and $N_\mathrm{TS}$ is presented in Tab~\ref{table3}. As we can see, in the vast majority of cases the proposed ASNet requires less time than GAS, while the proposed BFNet always executes faster than CDM-Alt or Babai-Alt. Generally, the proposed approach requires approximately $1\, \text{ms}$ to process each sample, implying possible applications in mmWave communications where the channel coherence time is in the order of milliseconds\cite{7742901}.}

\section{Conclusion}
In this paper, we designed a hybrid beamforming architecture with antenna selection to provide rich flexibility and high hardware efficiency. We proposed a deep unsupervised learning-based approach consisting of the antenna selection network (ASNet) and the hybrid beamforming network (BFNet) to effectively solve the joint antenna selection and hybrid beamforming problem. Both two networks employ ResNet for extracting features from the channel matrix. To enable unsupervised learning, we proposed a deep probabilistic subsampling trick for ASNet and a specially designed quantization function for BFNet, and thus the discrete problem constraints can be incorporated into the joint network while preserving differentiability. We proposed a flexible loss function, enabling a phased unsupervised training approach to train the joint network efficiently. Hence, we avoid the difficulty of acquiring training labels commonly faced by supervised learning. Simulation results demonstrate the outperformance of the proposed approach over conventional algorithms in terms of achieved rate and computational complexity. 

\ifCLASSOPTIONcaptionsoff
  \newpage
\fi

\bibliographystyle{IEEEtran}
\bibliography{IEEEabrv.bib, mybib.bib}

\begin{thebibliography}{10}
\providecommand{\url}[1]{#1}
\csname url@samestyle\endcsname
\providecommand{\newblock}{\relax}
\providecommand{\bibinfo}[2]{#2}
\providecommand{\BIBentrySTDinterwordspacing}{\spaceskip=0pt\relax}
\providecommand{\BIBentryALTinterwordstretchfactor}{4}
\providecommand{\BIBentryALTinterwordspacing}{\spaceskip=\fontdimen2\font plus
\BIBentryALTinterwordstretchfactor\fontdimen3\font minus
  \fontdimen4\font\relax}
\providecommand{\BIBforeignlanguage}[2]{{%
\expandafter\ifx\csname l@#1\endcsname\relax
\typeout{** WARNING: IEEEtran.bst: No hyphenation pattern has been}%
\typeout{** loaded for the language `#1'. Using the pattern for}%
\typeout{** the default language instead.}%
\else
\language=\csname l@#1\endcsname
\fi
#2}}
\providecommand{\BIBdecl}{\relax}
\BIBdecl

\bibitem{1}
F.~{Rusek}, D.~{Persson}, B.~K. {Lau}, E.~G. {Larsson}, T.~L. {Marzetta},
  O.~{Edfors}, and F.~{Tufvesson}, ``Scaling up mimo: Opportunities and
  challenges with very large arrays,'' \emph{{IEEE} Signal Processing Mag.},
  vol.~30, no.~1, pp. 40--60, 2013.

\bibitem{7010533}
S.~{Han}, C.~{I}, Z.~{Xu}, and C.~{Rowell}, ``Large-scale antenna systems with
  hybrid analog and digital beamforming for millimeter wave 5g,'' \emph{{IEEE}
  Commun. Mag.}, vol.~53, no.~1, pp. 186--194, 2015.

\bibitem{hbf1}
A.~{Alkhateeb}, J.~{Mo}, N.~{Gonzalez-Prelcic}, and R.~W. {Heath}, ``Mimo
  precoding and combining solutions for millimeter-wave systems,'' \emph{{IEEE}
  Commun. Mag.}, vol.~52, no.~12, pp. 122--131, 2014.

\bibitem{hbf2}
O.~E. {Ayach}, S.~{Rajagopal}, S.~{Abu-Surra}, Z.~{Pi}, and R.~W. {Heath},
  ``Spatially sparse precoding in millimeter wave mimo systems,'' \emph{{IEEE}
  Trans. Wireless Commun.}, vol.~13, no.~3, pp. 1499--1513, 2014.

\bibitem{moaltmin}
X.~{Yu}, J.~{Shen}, J.~{Zhang}, and K.~B. {Letaief}, ``Alternating minimization
  algorithms for hybrid precoding in millimeter wave mimo systems,''
  \emph{{IEEE} J. Sel. Topics Signal Process.}, vol.~10, no.~3, pp. 485--500,
  2016.

\bibitem{hbf3}
C.~{Rusu}, R.~{Mèndez-Rial}, N.~{González-Prelcic}, and R.~W. {Heath}, ``Low
  complexity hybrid precoding strategies for millimeter wave communication
  systems,'' \emph{{IEEE} Trans. Wireless Commun.}, vol.~15, no.~12, pp.
  8380--8393, 2016.

\bibitem{7370753}
R.~{Méndez-Rial}, C.~{Rusu}, N.~{González-Prelcic}, A.~{Alkhateeb}, and R.~W.
  {Heath}, ``Hybrid mimo architectures for millimeter wave communications:
  Phase shifters or switches?'' \emph{IEEE Access}, vol.~4, pp. 247--267, 2016.

\bibitem{sohrabi1}
F.~{Sohrabi} and W.~{Yu}, ``Hybrid beamforming with finite-resolution phase
  shifters for large-scale mimo systems,'' in \emph{Proc. IEEE 16th Int.
  Workshop Signal Process. Adv. Wireless Commun. (SPAWC)}, Stockholm, Sweden,
  2015, pp. 136--140.

\bibitem{sohrabi2}
------, ``Hybrid digital and analog beamforming design for large-scale antenna
  arrays,'' \emph{{IEEE} J. Sel. Topics Signal Process.}, vol.~10, no.~3, pp.
  501--513, 2016.

\bibitem{jchen}
J.~{Chen}, ``Hybrid beamforming with discrete phase shifters for
  millimeter-wave massive mimo systems,'' \emph{{IEEE} Trans. Vehicular
  Technol.}, vol.~66, no.~8, pp. 7604--7608, 2017.

\bibitem{zwang}
Z.~{Wang}, M.~{Li}, Q.~{Liu}, and A.~L. {Swindlehurst}, ``Hybrid precoder and
  combiner design with low-resolution phase shifters in mmwave mimo systems,''
  \emph{{IEEE} J. Sel. Topics Signal Process.}, vol.~12, no.~2, pp. 256--269,
  2018.

\bibitem{9411813}
S.~Lyu, Z.~Wang, Z.~Gao, H.~He, and L.~Hanzo, ``Lattice-based mmwave hybrid
  beamforming,'' \emph{{IEEE} Trans. Commun.}, vol.~69, no.~7, pp. 4907--4920,
  2021.

\bibitem{2}
Y.~{Gao}, H.~{Vinck}, and T.~{Kaiser}, ``Massive mimo antenna selection:
  Switching architectures, capacity bounds, and optimal antenna selection
  algorithms,'' \emph{{IEEE} Trans. Signal Processing}, vol.~66, no.~5, pp.
  1346--1360, 2018.

\bibitem{3}
A.~{Dua}, K.~{Medepalli}, and A.~J. {Paulraj}, ``Receive antenna selection in
  mimo systems using convex optimization,'' \emph{{IEEE} Trans. Wireless
  Commun.}, vol.~5, no.~9, pp. 2353--2357, 2006.

\bibitem{4}
B.~H. {Wang}, H.~T. {Hui}, and M.~S. {Leong}, ``Global and fast receiver
  antenna selection for mimo systems,'' \emph{{IEEE} Trans. Commun.}, vol.~58,
  no.~9, pp. 2505--2510, 2010.

\bibitem{5}
M.~{Gharavi-Alkhansari} and A.~B. {Gershman}, ``Fast antenna subset selection
  in mimo systems,'' \emph{{IEEE} Trans. Signal Processing}, vol.~52, no.~2,
  pp. 339--347, 2004.

\bibitem{6}
X.~{Zhai}, Y.~{Cai}, Q.~{Shi}, M.~{Zhao}, G.~Y. {Li}, and B.~{Champagne},
  ``Joint transceiver design with antenna selection for large-scale mu-mimo
  mmwave systems,'' \emph{{IEEE} J. Select. Areas Commun.}, vol.~35, no.~9, pp.
  2085--2096, 2017.

\bibitem{hli}
H.~{Li}, Q.~{Liu}, Z.~{Wang}, and M.~{Li}, ``Joint antenna selection and analog
  precoder design with low-resolution phase shifters,'' \emph{{IEEE} Trans.
  Vehicular Technol.}, vol.~68, no.~1, pp. 967--971, 2019.

\bibitem{7}
H.~{He}, C.~{Wen}, S.~{Jin}, and G.~Y. {Li}, ``Deep learning-based channel
  estimation for beamspace mmwave massive mimo systems,'' \emph{{IEEE} Wireless
  Commun. Lett.}, vol.~7, no.~5, pp. 852--855, 2018.

\bibitem{9}
Y.~S. {Nasir} and D.~{Guo}, ``Multi-agent deep reinforcement learning for
  dynamic power allocation in wireless networks,'' \emph{{IEEE} J. Select.
  Areas Commun.}, vol.~37, no.~10, pp. 2239--2250, 2019.

\bibitem{8}
H.~{Ye}, G.~Y. {Li}, and B.~{Juang}, ``Power of deep learning for channel
  estimation and signal detection in ofdm systems,'' \emph{{IEEE} Wireless
  Commun. Lett.}, vol.~7, no.~1, pp. 114--117, 2018.

\bibitem{8742579}
A.~Zappone, M.~Di~Renzo, and M.~Debbah, ``Wireless networks design in the era
  of deep learning: Model-based, ai-based, or both?'' \emph{{IEEE} Trans.
  Wireless Commun.}, vol.~67, no.~10, pp. 7331--7376, 2019.

\bibitem{8227766}
H.~Sun, X.~Chen, Q.~Shi, M.~Hong, X.~Fu, and N.~D. Sidiropoulos, ``Learning to
  optimize: Training deep neural networks for wireless resource management,''
  in \emph{Proc. IEEE 17th Int. Workshop Signal Process. Adv. Wireless Commun.
  (SPAWC)}, Sapporo, Japan, 2017, pp. 1--6.

\bibitem{mlhbf1}
A.~M. {Elbir}, ``Cnn-based precoder and combiner design in mmwave mimo
  systems,'' \emph{{IEEE} Commun. Lett.}, vol.~23, no.~7, pp. 1240--1243, 2019.

\bibitem{mlhbf2}
T.~{Peken}, S.~{Adiga}, R.~{Tandon}, and T.~{Bose}, ``Deep learning for svd and
  hybrid beamforming,'' \emph{{IEEE} Trans. Wireless Commun.}, vol.~19, no.~10,
  pp. 6621--6642, 2020.

\bibitem{mlhbf3}
Q.~{Wang}, K.~{Feng}, X.~{Li}, and S.~{Jin}, ``Precodernet: Hybrid beamforming
  for millimeter wave systems with deep reinforcement learning,'' \emph{{IEEE}
  Wireless Commun. Lett.}, vol.~9, no.~10, pp. 1677--1681, 2020.

\bibitem{mlhbf4}
H.~{He}, M.~{Zhang}, S.~{Jin}, C.~K. {Wen}, and G.~Y. {Li}, ``Model-driven deep
  learning for massive mu-mimo with finite-alphabet precoding,'' \emph{{IEEE}
  Commun. Lett.}, vol.~24, no.~10, pp. 2216--2220, 2020.

\bibitem{mlas2}
Y.~{Yang}, S.~{Zhang}, F.~{Gao}, C.~{Xu}, J.~{Ma}, and O.~A. {Dobre}, ``Deep
  learning based antenna selection for channel extrapolation in fdd massive
  mimo,'' in \emph{Proc. Int. Conf. Wireless Commun. Signal Process. (WCSP)},
  Nanjing, China, 2020, pp. 182--187.

\bibitem{mlas3}
B.~Lin, F.~Gao, S.~Zhang, T.~Zhou, and A.~Alkhateeb, ``Deep learning based
  antenna selection and csi extrapolation in massive mimo systems,''
  \emph{{IEEE} Trans. Wireless Commun.}, vol.~20, no.~11, pp. 7669--7681, 2021.

\bibitem{10}
J.~{Joung}, ``Machine learning-based antenna selection in wireless
  communications,'' \emph{{IEEE} Commun. Lett.}, vol.~20, no.~11, pp.
  2241--2244, 2016.

\bibitem{mlas1}
M.~S. {Ibrahim}, A.~S. {Zamzam}, X.~{Fu}, and N.~D. {Sidiropoulos},
  ``Learning-based antenna selection for multicasting,'' in \emph{Proc. IEEE
  19th Int. Workshop Signal Process. Adv. Wireless Commun. (SPAWC)}, Kalamata,
  Greece, 2018, pp. 1--5.

\bibitem{11}
A.~M. {Elbir} and K.~V. {Mishra}, ``Joint antenna selection and hybrid
  beamformer design using unquantized and quantized deep learning networks,''
  \emph{{IEEE} Trans. Wireless Commun.}, vol.~19, no.~3, pp. 1677--1688, 2020.

\bibitem{tvu}
T.~X. Vu, S.~Chatzinotas, V.-D. Nguyen, D.~T. Hoang, D.~N. Nguyen, M.~D. Renzo,
  and B.~Ottersten, ``Machine learning-enabled joint antenna selection and
  precoding design: From offline complexity to online performance,''
  \emph{{IEEE} Trans. Wireless Commun.}, vol.~20, no.~6, pp. 3710--3722, 2021.

\bibitem{unsup1}
J.~{Xu}, P.~{Zhu}, J.~{Li}, and X.~{You}, ``Deep learning-based pilot design
  for multi-user distributed massive mimo systems,'' \emph{{IEEE} Wireless
  Commun. Lett.}, vol.~8, no.~4, pp. 1016--1019, 2019.

\bibitem{unsup2}
H.~{Huang}, W.~{Xia}, J.~{Xiong}, J.~{Yang}, G.~{Zheng}, and X.~{Zhu},
  ``Unsupervised learning-based fast beamforming design for downlink mimo,''
  \emph{IEEE Access}, vol.~7, pp. 7599--7605, 2019.

\bibitem{unsup4}
T.~{Lin} and Y.~{Zhu}, ``Beamforming design for large-scale antenna arrays
  using deep learning,'' \emph{{IEEE} Wireless Commun. Lett.}, vol.~9, no.~1,
  pp. 103--107, 2020.

\bibitem{unsuphb3}
H.~Hojatian, J.~Nadal, J.-F. Frigon, and F.~Leduc-Primeau, ``Unsupervised deep
  learning for massive mimo hybrid beamforming,'' \emph{{IEEE} Trans. Wireless
  Commun.}, vol.~20, no.~11, pp. 7086--7099, 2021.

\bibitem{15}
Y.~{Gao} and T.~{Kaiser}, ``Antenna selection in massive mimo systems:
  Full-array selection or subarray selection?'' in \emph{Proc. IEEE Sensor
  Array Multichannel Signal Process. Workshop}, Rio de Janeiro, Brazil, 2016,
  pp. 1--5.

\bibitem{pilotpc1}
F.~Sohrabi, K.~M. Attiah, and W.~Yu, ``Deep learning for distributed channel
  feedback and multiuser precoding in fdd massive mimo,'' \emph{{IEEE} Trans.
  Wireless Commun.}, vol.~20, no.~7, pp. 4044--4057, 2021.

\bibitem{pilotpc2}
K.~M. Attiah, F.~Sohrabi, and W.~Yu, ``Deep learning approach to channel
  sensing and hybrid precoding for tdd massive mimo systems,'' in \emph{Proc.
  IEEE GLOBECOM}, Taipei, Taiwan, 2020, pp. 1--6.

\bibitem{goodfellow2016deep}
I.~Goodfellow, Y.~Bengio, A.~Courville, and Y.~Bengio, \emph{Deep
  learning}.\hskip 1em plus 0.5em minus 0.4em\relax MIT press Cambridge, 2016,
  vol.~1, no.~2.

\bibitem{resnet}
K.~He, X.~Zhang, S.~Ren, and J.~Sun, ``Deep residual learning for image
  recognition,'' in \emph{Proc. IEEE Conf. Comput. Vis. Pattern Recognit.
  (CVPR)}, Las Vegas, NV, USA, June 2016.

\bibitem{resnet1}
Y.~Cao, T.~Lv, Z.~Lin, P.~Huang, and F.~Lin, ``Complex resnet aided doa
  estimation for near-field mimo systems,'' \emph{{IEEE} Trans. Vehicular
  Technol.}, vol.~69, no.~10, pp. 11\,139--11\,151, 2020.

\bibitem{resnet2}
Y.~Tian, G.~Pan, and M.-S. Alouini, ``Applying deep-learning-based computer
  vision to wireless communications: Methodologies, opportunities, and
  challenges,'' \emph{IEEE Open J. Commun. Soc.}, vol.~2, pp. 132--143, 2021.

\bibitem{pmlr-v37-ioffe15}
S.~Ioffe and C.~Szegedy, ``Batch normalization: Accelerating deep network
  training by reducing internal covariate shift,'' in \emph{Proc. of Intl.
  Conf. on Machine Learning (ICML)}, vol.~37, Lille, France, July 2015, pp.
  448--456.

\bibitem{huijben2019deep}
I.~A. Huijben, B.~S. Veeling, and R.~J. van Sloun, ``Deep probabilistic
  subsampling for task-adaptive compressed sensing,'' in \emph{Proc. Int. Conf.
  Learning Rep.}, Addis Ababa, Ethiopia, 2020.

\bibitem{gumbel1954statistical}
E.~J. Gumbel, \emph{Statistical theory of extreme values and some practical
  applications: a series of lectures}.\hskip 1em plus 0.5em minus 0.4em\relax
  US Government Printing Office, 1954, vol.~33.

\bibitem{18}
E.~Jang, S.~Gu, and B.~Poole, ``Categorical reparameterization with
  gumbel-softmax,'' \emph{arXiv:1611.01144}, 2016.

\bibitem{19}
C.~J. Maddison, A.~Mnih, and Y.~W. Teh, ``The concrete distribution: A
  continuous relaxation of discrete random variables,''
  \emph{arXiv:1611.00712}, 2016.

\bibitem{l2}
A.~Y. Ng, ``Feature selection, l1 vs. l2 regularization, and rotational
  invariance,'' in \emph{Proc. of Intl. Conf. on Machine Learning (ICML)},
  Banff, Alberta, Canada, 2004, p.~78.

\bibitem{adam}
D.~P. Kingma and J.~Ba, ``Adam: A method for stochastic optimization,''
  \emph{arXiv:1412.6980}, 2014.

\bibitem{cnntimecost}
K.~He and J.~Sun, ``Convolutional neural networks at constrained time cost,''
  in \emph{Proc. IEEE Conf. Comput. Vis. Pattern Recognit. (CVPR)}, Boston, MA,
  USA, June 2015.

\bibitem{dnnconv}
Z.~Allen-Zhu, Y.~Li, and Z.~Song, ``A convergence theory for deep learning via
  over-parameterization,'' in \emph{Proc. of Intl. Conf. on Machine Learning
  (ICML)}, vol.~97, Long Beach, CA, USA, June 2019, pp. 242--252.

\bibitem{wi}
Remcom, ``Wireless insite,'' [Online]. Available:
  https://www.remcom.com/wireless-insite-em-propagation-software.

\bibitem{lmmse}
M.~Biguesh and A.~Gershman, ``Training-based mimo channel estimation: a study
  of estimator tradeoffs and optimal training signals,'' \emph{{IEEE} Trans.
  Signal Processing}, vol.~54, no.~3, pp. 884--893, 2006.

\bibitem{4411574}
R.~Chen, J.~G. Andrews, and R.~W. Heath, ``Efficient transmit antenna selection
  for multiuser mimo systems with block diagonalization,'' in \emph{Proc. IEEE
  GLOBECOM}, Washington, D.C., USA, 2007, pp. 3499--3503.

\bibitem{7742901}
V.~Va, J.~Choi, and R.~W. Heath, ``The impact of beamwidth on temporal channel
  variation in vehicular channels and its implications,'' \emph{{IEEE} Trans.
  Vehicular Technol.}, vol.~66, no.~6, pp. 5014--5029, 2017.

\end{thebibliography}

\begin{IEEEbiography}[{\includegraphics[width=1in,height=1.25in,clip,keepaspectratio]{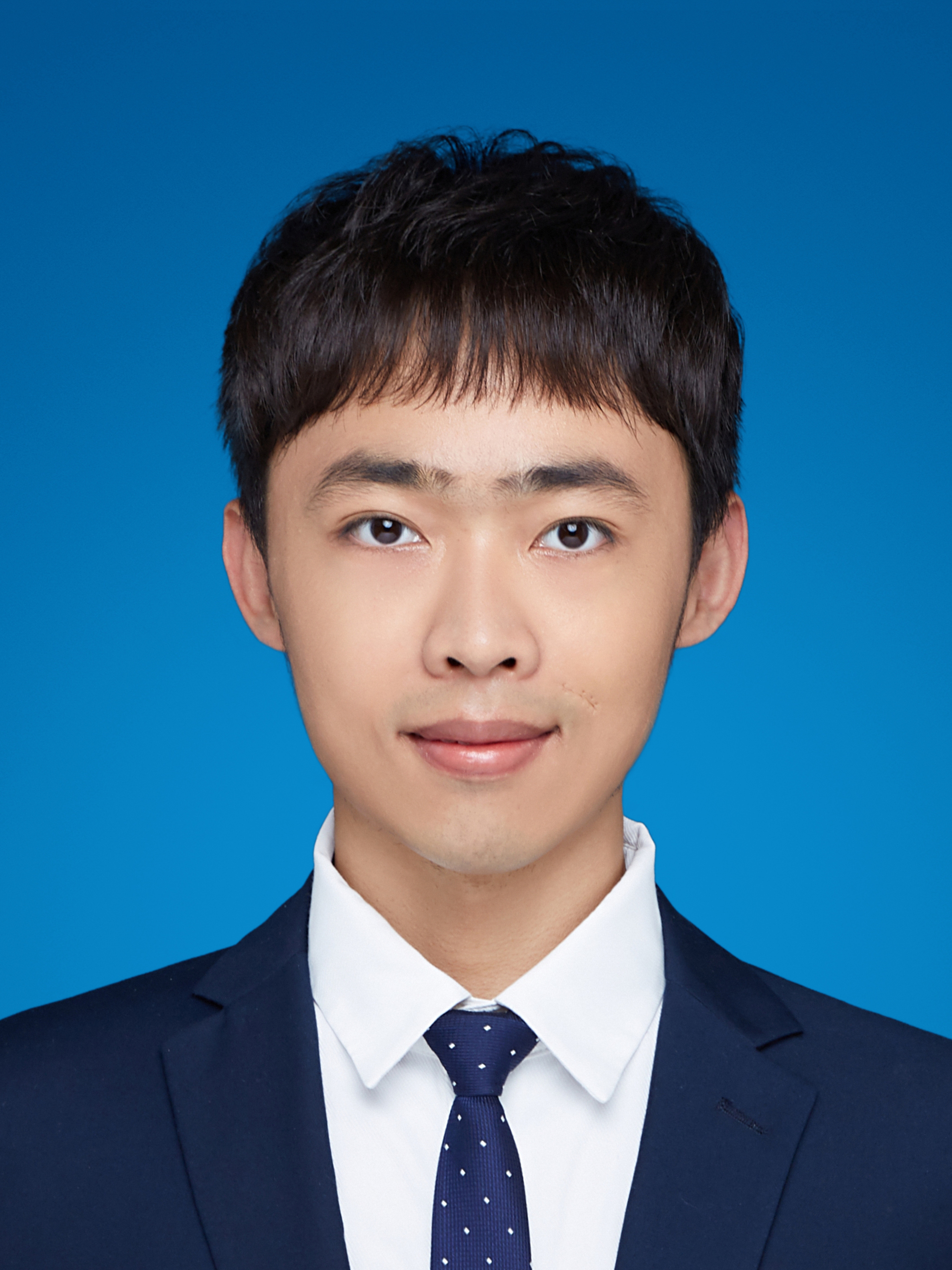}}]{Zhiyan Liu} received the B.Eng. degree in 2021 from the Department of Electronic Engineering, Tsinghua University. He is currently pursuing the Ph.D. degree with the Department of Electrical and Electronic Engineering, The University of Hong Kong (HKU). 
  His research interests include massive multiple-input-multiple-output (MIMO)
  and edge intelligence.
  \end{IEEEbiography}

  \begin{IEEEbiography}[{\includegraphics[width=1in,height=1.25in,clip,keepaspectratio]{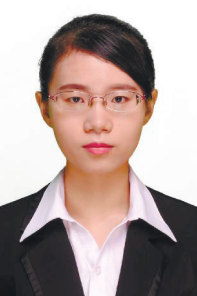}}]{Yuwen Yang} (S'20) received B.S. degree in Telecommunication Engineering from Xidian University, China, in 2018. She is currently working towards the Ph.D. degree with
  the Department of Automation, Tsinghua University, Beijing, China.
  Her research interests include massive multiple-input-multiple-output (MIMO)
  and machine learning for wireless communications.
  \end{IEEEbiography}

  \begin{IEEEbiography}[{\includegraphics[width=1in,height=1.25in,clip,keepaspectratio]{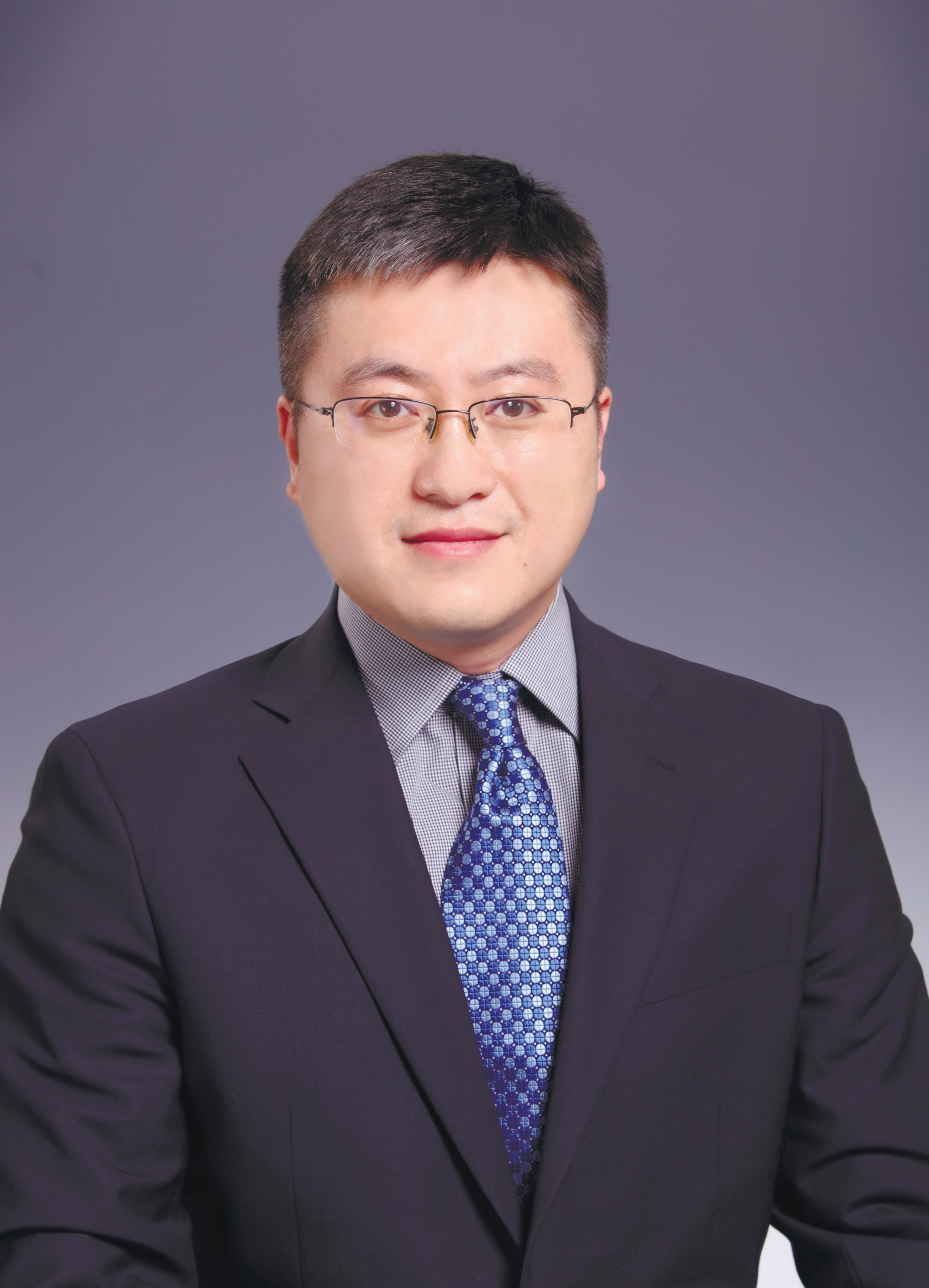}}]
  {Feifei Gao} (M'09-SM'14-F'20) received the B.Eng. degree from Xi'an Jiaotong University, Xi'an, China in 2002, the M.Sc. degree from McMaster University, Hamilton, ON, Canada in 2004, and the Ph.D. degree from National University of Singapore, Singapore in 2007. Since 2011, he joined the Department of Automation, Tsinghua University, Beijing, China, where he is currently an Associate Professor.
  
  Prof. Gao's research interests include signal processing for communications, array signal processing, convex optimizations, and artificial intelligence assisted communications. He has authored/coauthored more than 150 refereed IEEE journal papers and more than 150 IEEE conference proceeding papers that are cited more than 11800 times in Google Scholar. Prof. Gao has served as an Editor of IEEE Transactions on Wireless Communications, IEEE Journal of Selected Topics in Signal Processing (Lead Guest Editor), IEEE Transactions on Cognitive Communications and Networking, IEEE Signal Processing Letters (Senior Editor), IEEE Communications Letters (Senior Editor), IEEE Wireless Communications Letters, and China Communications. He has also served as the symposium co-chair for 2019 IEEE Conference on Communications (ICC), 2018 IEEE Vehicular Technology Conference Spring (VTC), 2015 IEEE Conference on Communications (ICC), 2014 IEEE Global Communications Conference (GLOBECOM), 2014 IEEE Vehicular Technology Conference Fall (VTC), as well as Technical Committee Members for more than 50 IEEE conferences.
  \end{IEEEbiography}

  \begin{IEEEbiography}[{\includegraphics[width=1in,height=1.25in,clip,keepaspectratio]{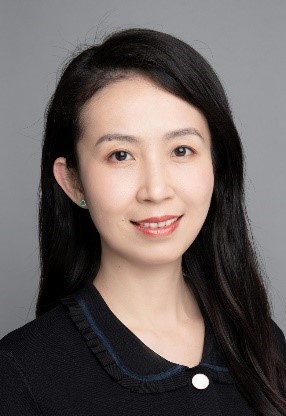}}]
  {Ting Zhou} received the B.S. and M.S. degrees from the Department of Electronic Engineering, Tsinghua University, in 2004 and 2006, respectively, and the Ph.D. degree from the Shanghai Institute of Microsystem and Information Technology (SIMIT), Chinese Academy of Sciences, in 2011. She is now a Professor with Shanghai Advanced Research Institute, CAS. She has published over 70 journal and conference papers, and holds more than 60 granted or filed patents. Her research interests include wireless resource management and mobility management, intelligent networking of heterogeneous wireless networks, and 6G mobile systems. She has won the 2020 First Prize of Shanghai Science and Technology Award and the 2019, 2016 and 2015 First Prize of Technological Invention Awards from China Institute of Communications.
  \end{IEEEbiography}

  \begin{IEEEbiography}[{\includegraphics[width=1in,height=1.25in,clip,keepaspectratio]{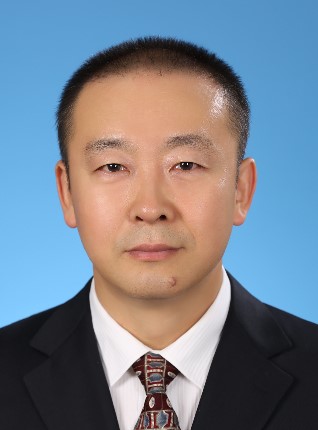}}]
  {Hongbing Ma} received the PhD degree from Peking University, China in 1999. He is currently a professor with the Department of Electronic Engineering, Tsinghua University, China. His research interests include image processing, pattern recognition, and spatial information processing and application.
  \end{IEEEbiography}

\end{document}